\documentclass[twocolumn]{aastex701}

\newcommand{\deltags}{{\delta}_{\mathrm{g}}^s}

\usepackage{amsmath}
\begin{document}

\title{Reconstructing the Large-Scale Velocity Field from DESI DR1 with Bayesian Inference}

\author[0009-0006-4297-2678]{Raeed Mundow}
\email{raeed7sm@gmail.com}
\affiliation{ Department of Physics and the Asher Space Research Institute\\ Israel Institute of Technology Technion, Haifa 32000, Israel}

\author[0000-0002-8272-4779]{Adi Nusser}   
\email{adi@physics.technion.ac.il}
\affiliation{ Department of Physics and the Asher Space Research Institute\\ Israel Institute of Technology Technion, Haifa 32000, Israel}
\begin{abstract}
Survey boundaries and nonlinear dynamics complicate reconstruction of
large-scale matter-density and peculiar-velocity fields from galaxy redshift
surveys. We develop and compare two reconstruction methods within a common
Bayesian framework, based on linear perturbation theory and the Zel'dovich
approximation, and apply them to the DESI Data Release 1 Bright Galaxy Survey.
We infer the initial density field using a Gaussian prior and likelihood,
with the Zel'dovich model displacing particles from a uniform Lagrangian
grid. Both yield maximum a posteriori (MAP) fields; only the linear
posterior is sampled.
Our main findings are:
\textit{(i)} Both reconstructions yield similar large-scale density structures
and coherent velocity patterns in the survey interior, beyond roughly
\(50\,h^{-1}\,\mathrm{Mpc}\) from the footprint boundaries.
\textit{(ii)} Comparison with Cosmicflows-4 groups, predominantly from the
SDSS Fundamental Plane sample in the DESI overlap, shows close agreement
between mean observed radial velocities and MAP predictions when binned
by the latter, despite substantial individual scatter.
\textit{(iii)} Both yield similar reflex-dipole profiles, defined as mean
shell velocities relative to those at randomly sampled centres, out to
\(\sim225\,h^{-1}\,\mathrm{Mpc}\). The profiles depend weakly on the prior
normalization \(\sigma_8\), but lie below the fiducial linear
\(\Lambda\)CDM reference at \(r<50\,h^{-1}\,\mathrm{Mpc}\).
Sparse sampling limits the largest radii.
\textit{(iv)} In an exploratory BAO test, shifting galaxies and randoms
to a target redshift \(z=8\) using reconstructed displacements sharpens
the acoustic feature in both cases, with a slightly more pronounced
peak for Zel'dovich.
Survey geometry and galaxy bias remain the principal limitations.
\end{abstract}


\section{Introduction} \label{sec:intro}

Peculiar motions of matter leave signatures in a wide range of observations
of large-scale structure, making their effects essential to the interpretation
of galaxy surveys. Peculiar velocities contribute to the observed galaxy
redshifts and hence shift the inferred line-of-sight positions, distorting
the apparent clustering pattern in redshift space \citep{Kaiser1987}.
The same perturbations to redshift-based distances modulate the absolute
luminosities inferred from observed fluxes. This effect was used to constrain
the relative motion of the Virgo Cluster and the Local Group
\citep{Tammann1979} and was later generalized into a statistical method for
measuring coherent flows in large redshift surveys
\citep{Nusser2011Luminosities}. Peculiar velocities also provide a direct
dynamical probe of the three-dimensional total matter distribution, which is
dominated by dark matter \citep{StraussWillick1995}. By the equivalence
principle, freely falling tracers respond to the same gravitational
acceleration irrespective of their composition. On sufficiently large scales,
where nongravitational baryonic effects are negligible, galaxies are therefore
expected to trace the bulk velocity field of the underlying matter. This
makes peculiar velocities especially valuable for probing the matter
distribution without requiring galaxy number density to trace mass directly.

Extracting this dynamical information from peculiar-velocity observations is,
however, challenging. Direct measurements compare an observed redshift with
a redshift-independent distance estimate obtained from the Tully--Fisher
relation, the Fundamental Plane, or the standardized luminosity of a Type Ia
supernova. Such measurements have revealed coherent flows in the nearby
Universe, including the streaming motion associated with the Great Attractor
\citep{LyndenBell1988}. The resulting catalogs nevertheless sample the
velocity field sparsely and unevenly compared with galaxy redshift surveys,
and constrain only its line-of-sight component. Recovering the full
three-dimensional flow requires additional modeling assumptions.
Furthermore, the scatter in distance indicators translates into velocity
uncertainties that grow with distance, limiting the depth over which
individual measurements are informative. Distance errors coupled to survey
selection and inhomogeneous galaxy sampling can introduce Malmquist and
selection biases, while uncertainties in distance calibration can generate
systematic velocity offsets
\citep{StraussWillick1995,Nusser2026Distance}.

Galaxy redshift surveys provide a complementary route to the peculiar-velocity
field through gravitational instability. Matter-density fluctuations generate
peculiar gravitational accelerations, which drive departures from the uniform
Hubble expansion. On sufficiently large scales, the growing-mode velocity is
proportional to the gravitational acceleration, with a proportionality factor
determined by the background expansion and the growth of structure
\citep{Peebles1980}. Thus, given a cosmological model and a prescription for
galaxy bias, the observed galaxy distribution can be used to predict coherent
three-dimensional peculiar motions \citep{1994ApJ...421L...1N}.

This density-based approach benefits from the dense sampling of redshift
surveys and does not require independent distance measurements for individual
galaxies. Several effects must nevertheless be accounted for. Galaxies are
\textit{biased} tracers of the underlying matter distribution, so a bias
prescription is needed to relate their density to \(\delta_m\). Moreover,
surveys provide  redshifts rather than true distances: the observed galaxy
distribution is distorted by the very velocities being inferred, requiring
a consistent treatment of redshift-space distortions within the adopted
dynamical model. The angular footprint and magnitude limits further restrict
the observed volume and produce spatially varying sampling and shot noise.
Because the velocity field depends on matter beyond the observed region,
survey boundaries and unobserved structure must also be treated in the
reconstruction.

Several approaches have been developed to address these challenges. A
straightforward strategy is to smooth the weighted galaxy distribution,
producing a continuous density field that can be evaluated on a grid.
A sufficiently broad smoothing window reduces shot noise and suppresses
the contribution of small-scale nonlinear structure, restricting the
velocity prediction to scales where linear theory is an adequate
approximation. This strategy underlies early IRAS reconstructions
\citep{Yahil1991,Davis1991}, the redshift-space treatment of
\citet{1994ApJ...421L...1N}, and later analyses of 2MRS and 2M++
\citep{Davis2011,Carrick2015}. Redshift-space distortions are accounted
for through iterative corrections to galaxy positions or a direct solution
in redshift space. The smoothing scale sets a trade-off between noise
suppression and spatial resolution. These methods have the advantage that the smoothing scale is
precisely controlled, allowing comparison with directly measured
peculiar-velocity fields smoothed on an equivalent scale.

Recent tests of this approach for DESI and Rubin/LSST use mock galaxy
catalogs drawn from the AbacusSummit simulations.
\citet{2024PhRvD.109j3533R} study DESI-like luminous red galaxies in a
periodic box at \(z=0.5\), while \citet{2024PhRvD.109j3534H} use
light-cone mocks with evolving tracer populations and realistic survey
selection and geometry. Both employ BAO-style reconstruction with the
\texttt{pyrecon} multigrid solver: a smoothed galaxy-density field and an
assumed linear bias enter the linearized continuity equation in redshift
space, yielding displacements that are converted to velocities through
\(\mathbf{v}=aHf\boldsymbol{\Psi}\). These studies assess velocity
recovery against the simulated velocities, principally to calibrate
applications to the kinematic Sunyaev--Zel'dovich effect.

Alternatively, Bayesian approaches combine a statistical prior for the
underlying fields with a likelihood relating those fields to the survey
observations. With a linear response and Gaussian prior and likelihood,
Wiener filtering gives the posterior mean, while constrained realizations
sample the corresponding field uncertainty
\citep{10.1093/mnras/272.4.885,1991ApJ...380L...5H,1999ApJ...520..413Z}.
Bayesian forward modeling also allows nonlinear dynamical approximations
to be incorporated into the inference. Applications to 2MRS include
Wiener-filtered density and velocity fields with constrained realizations
within \(200\,h^{-1}\,\mathrm{Mpc}\) \citep{LilowNusser2021} and, more
recently, a Zel'dovich forward model combining an unbinned likelihood with
Hamiltonian Monte Carlo sampling and comparison with Cosmicflows-4
\citep{2026arXiv260608593N}. These applications provide detailed views of
the local flow, but probe relatively small volumes and predominantly very
low redshift.

In this paper, we reconstruct the matter-density and peculiar-velocity
fields from the DESI DR1 Bright Galaxy Survey out to
\(z\lesssim0.5\), covering a substantially larger volume
than those probed by 2MRS and 2M++. The fields are represented in a
\(3\,h^{-1}\,\mathrm{Gpc}\) computational box. Within a common Bayesian
framework (Section~\ref{sec:bayesian_framework}), we infer an initial
density field whose evolution is described by either linear perturbation
theory or the Zel'dovich approximation. For each dynamical approximation,
the forward model predicts the observed redshift-space galaxy distribution,
incorporating tracer bias, evolution along the light cone, redshift-space
distortions, and the DESI angular mask and radial selection. Both models
use a Gaussian prior for the initial density field and a Gaussian
likelihood for the observed galaxy overdensity, yielding MAP
reconstructions of the density and velocity fields. The distinction from
the direct continuity-equation reconstructions above is the inference of
a latent initial field through an explicit prior and survey likelihood.
In particular, our Zel'dovich model evolves particles to predict the
redshift-space density entering the likelihood. It does not simply convert
BAO-style displacements estimated from the smoothed observed density into
velocities.

Our analysis has three main objectives. First, we assess the recovery of the
known velocity field in controlled mocks. Second, we examine how survey
boundaries, the selection function, and the cosmological prior affect the
DESI reconstructions. Third, we test gravitational instability  by comparing the peculiar velocities
predicted from the DESI density distribution with the observed radial
peculiar velocities of Cosmicflows-4 (CF4) groups in the overlapping volume.
These distance-based velocities are not used to constrain either
reconstruction. The comparison therefore tests whether the gravitational
growth of the matter distribution traced by DESI accounts for the
independently observed coherent motions, under the adopted tracer-bias and
survey model. We also present an exploratory application of the
reconstructed displacement fields to BAO reconstruction
\citep{Eisenstein_2007,2015MNRAS.453..456B}.

Throughout the analysis, we adopt a spatially flat Planck-2018
\(\Lambda\)CDM cosmology with \(h=0.6766\), \(\Omega_{\rm m}=0.3111\),
\(\Omega_{\rm b}=0.049\), \(n_{\rm s}=0.9665\), and a fiducial
normalization \(\sigma_8=0.81\), based on the TT,TE,EE+lowE+lensing+BAO results of
\citet{Planck2018}. Its linear matter power spectrum,
computed using \texttt{CAMB}
\citep{2000ApJ...538..473L,2026arXiv260714854L}, defines the Gaussian
cosmological prior shared by both reconstructions. The finite-grid
representation of this prior is described in Section~\ref{sec:grid}.

The paper is organized as follows. Section~\ref{sec:data} describes the DESI
and CF4 data sets. Section~\ref{sec:bayesian_framework} describes the
Bayesian reconstruction framework: the prior, the forward models based on
linear and Zel'dovich dynamics, the survey likelihood, and posterior
inference.
Section~\ref{sec:recon_setup} specifies the numerical setup and catalog
preparation, with technical gridding details in the appendix.
Section~\ref{sec:mock} assesses the resulting reconstructions using controlled MDPL2 mocks.
Section~\ref{sec:desi_results} presents the DESI MAP reconstructions and
boundary effects, linear constrained realizations, density and flow
diagnostics, the CF4 comparison, and an exploratory BAO application.
Section~\ref{sec:discussion} discusses the findings and limitations.
The appendices provide derivations and implementation details.

\section{Data}
\label{sec:data}

\subsection{DESI DR1 Bright Galaxy Survey}
\label{sec:data_desi}

We reconstruct the density and velocity fields from the DESI Data Release 1
Bright Galaxy Survey (BGS) \citep{2025arXiv250314745D}. DESI is a five-year
spectroscopic survey with a planned footprint of approximately
\(14{,}000\,\mathrm{deg}^2\). The BGS is observed during bright time and
comprises two samples: BGS BRIGHT, selected primarily with \(r<19.5\), and
BGS FAINT, extending to \(r<20.175\) with additional selection criteria.
We combine the North and South Galactic Cap catalogs out to
\(z\lesssim0.5\), corresponding to a comoving distance of approximately
\(1300\,h^{-1}\,\mathrm{Mpc}\). The resulting reconstruction catalog
contains approximately \(5.5\times10^6\) tracers.

The associated DESI random catalogs describe the angular footprint, radial
selection, and observational completeness without containing cosmological
clustering. We use 18 random catalogs for each Galactic cap, containing
approximately \(336\times10^6\) random points in total. The data and random
catalogs carry the supplied observational weights. Section~\ref{sec:recon_setup}
describes how these catalogs are converted to a gridded overdensity, noise
model, and survey mask.

\subsection{Cosmicflows-4 peculiar velocities}
\label{sec:data_cf4}
We use the grouped Cosmicflows-4 (CF4) catalog \citep{Tully2023} to test
the gravitational instability picture within the adopted \(\Lambda\)CDM
cosmological model. In this picture, the matter distribution traced by DESI
BGS and the peculiar motions measured by CF4 arise from the same
gravitational growth of structure. We compare the line-of-sight velocities
predicted from the DESI density field at the positions of CF4 groups with
their observed radial peculiar velocities, taken from the catalog's
\texttt{Vpds} field. Since CF4 is not included in
either reconstruction likelihood, it provides an independent observational
test of this dynamical connection between density and velocity.

CF4 combines redshift-independent distances from several indicators,
principally the Tully--Fisher and Fundamental Plane relations, with additional
supernova and local distance calibrators. The component that dominates our
selected overlap with DESI BGS is the SDSS Fundamental Plane sample
\citep{2022MNRAS.515..953H}, from which most of the selected CF4 groups are
drawn. The published parent CF4 compilation contains distances for
\(55{,}877\) galaxies assembled into \(38{,}065\) groups.

To mitigate inhomogeneous Malmquist bias CF4 groups are positioned at their redshifts rather than observed distances. 
 The selected
CF4 groups are shown with the DESI tracers in
Figure~\ref{fig:desi_cf4_tracers}. The spatial selection defining the overlap
sample is described in Section~\ref{sec:cf4_comparison}.

\section{Bayesian reconstruction framework}
\label{sec:bayesian_framework}

Our Bayesian framework aims to recover the underlying matter-density and
peculiar-velocity fields from the observed redshift-space distribution of
galaxies. We formulate the inference in terms of the initial density field
\(\delta_{\rm init}\), normalized to the present epoch by adopting
\(D(a=1)=1\). A Gaussian prior \(p(\delta_{\rm init})\), specified by the
fiducial linear matter power spectrum, describes the expected amplitudes
and spatial correlations of this field before conditioning on the data.

For each proposed \(\delta_{\rm init}\), a dynamical model predicts the
matter density and peculiar velocity at the epochs sampled by the survey.
We perform separate reconstructions of the same data using linear
perturbation theory and the Zel'dovich approximation, described in the
subsections below. To compare these predictions with the observations,
each forward model must also account for galaxy bias,
velocity-induced redshift-space distortions, the angular mask, and the
radial selection function. We denote the resulting prediction for the
gridded galaxy overdensity by \(\mathcal F[\delta_{\rm init}]\), and the
observed overdensity by \(d\).

The likelihood \(p(d\mid\mathcal F[\delta_{\rm init}])\) measures agreement
between this prediction and the data. Combining it with the prior gives
the posterior
\begin{equation}
    p(\delta_{\rm init}\mid d)
    \propto p\!\left(d\mid\mathcal F[\delta_{\rm init}]\right)\,
    p(\delta_{\rm init}).
    \label{eq:general_posterior}
\end{equation}
In the MAP estimate, the prior regularizes poorly
constrained modes by suppressing amplitudes unsupported by the data.
Posterior samples retain fluctuations in these modes, with statistics
approaching those of the prior as the observational constraints weaken,
while preserving correlations with the field constrained by the data.
The remainder of this section follows the steps of the inference: the
prior in Section~\ref{sec:prior_whitening}, the forward models in
Section~\ref{sec:forward_models}, the likelihood in
Section~\ref{sec:survey_response}, and posterior inference in
Section~\ref{sec:posterior_inference}. Each forward model is presented
alongside its dynamical approximation, from the initial field to the
predicted galaxy distribution.

\subsection{Gaussian prior and whitened fields}
\label{sec:prior_whitening}

On the reconstruction grid, \(\delta_{\rm init}\) is a vector with a
multivariate Gaussian distribution of zero mean and covariance
\(C_{\rm init}\):
\begin{equation}
    p(\delta_{\rm init})\propto
    \exp\!\left[-\frac{1}{2}\delta_{\rm init}^{\top}
    C_{\rm init}^{-1}\delta_{\rm init}\right].
    \label{eq:initial_density_prior}
\end{equation}
Statistical isotropy and homogeneity makes this covariance diagonal in Fourier space,
 set by the linear matter power spectrum
\(P(k)\) at the present-day normalization. 

We can equivalently express the initial field in terms of whitened Fourier
variables \(\xi\):
\begin{equation}
    \delta_{\rm init}(\mathbf{k})=\sqrt{P(k)}\,\xi(\mathbf{k}),
    \label{eq:initial_density_whitening}
\end{equation}
with the normalization factors appropriate to the discrete Fourier
convention. Whitening removes the scale-dependent prior
amplitude of each mode. The independent real degrees of freedom in
\(\xi\) then have unit Gaussian priors, so
\(p(\xi)\propto\exp(-\xi^{\top}\xi/2)\). Fourier modes at
\(\mathbf{k}\) and \(-\mathbf{k}\) are complex conjugates, ensuring a
real density field.

For fixed \(P(k)\), whitening is a linear change of variables that preserves
the prior on \(\delta_{\rm init}\). We can therefore perform the inference
in either representation and transform the result accordingly.

\subsection{Forward models}
\label{sec:forward_models}
\label{sec:dynamical_approximations}

The forward model \(\mathcal F\) maps a trial initial density field into
the observable galaxy distribution. It combines dynamical evolution with
an observation model: the evolved matter density is related to the galaxy
density through tracer bias, and the predicted velocities determine the
mapping to redshift space. We first specify the observational ingredients
shared by both models, then construct each prediction using its chosen
dynamical approximation.

To first order in peculiar velocity divided by the speed of light, the
radial position inferred from redshift is shifted according to
\begin{equation}
    \mathbf{s}=\mathbf{x}
    +\frac{v_r(\mathbf{x})}{aH}\,\hat{\mathbf{x}},
    \qquad
    v_r=\mathbf{u}\cdot\hat{\mathbf{x}},
    \label{eq:forward_rsd_mapping}
\end{equation}
where \(\mathbf{x}\) and \(\mathbf{s}\) are real-space and redshift-space
comoving positions, \(\mathbf{u}\) is the physical peculiar velocity,
and \(v_r\) is its radial component. The scale factor \(a\) and Hubble
rate \(H\) are evaluated at the relevant epoch. Spatial
variation in these shifts changes the tracer density through compression
or dilation of volume elements. These redshift-space distortions are
therefore part of the forward prediction for every proposed initial field.
The observed galaxy positions are kept in redshift space throughout the
inference.

Both predictions are evaluated along the survey light cone and on the
same observational grid as \(d\). A binary mask \(M\) selects the cells
within the observed footprint. The radial selection and completeness set
the expected tracer counts and noise weights through the random catalog.
The likelihood in Section~\ref{sec:survey_response} compares these masked
predictions with the data. Numerical choices for the grid, mask, and
catalog estimator are given in Section~\ref{sec:recon_setup}.

\subsubsection{Linear dynamics and forward model}
\label{sec:linear_dynamics}
\label{sec:linear_reconstruction}

In linear theory, the growing density perturbations evolve without changing
their spatial form \citep{Peebles1980}:
\begin{equation}
    \delta_m(\mathbf{x},a)=D(a)\,\delta_{\rm init}(\mathbf{x}),
    \label{eq:linear_density_growth}
\end{equation}
where \(\delta_m=(\rho_m-\bar\rho_m)/\bar\rho_m\), \(\mathbf{x}\) is
comoving position, and \(a\) is the scale factor. The linear growth factor
\(D(a)\) has the present-day normalization specified above. Its logarithmic
growth rate is
\(f(a)\equiv{\rm d}\ln D/{\rm d}\ln a=\dot D/(HD)\), where
\(H=\dot a/a\) and dots denote derivatives with respect to cosmic time.
A useful approximation is \(f(a)\simeq\Omega_m(a)^\gamma\), with
\(\gamma\simeq0.55\) for general relativity with a \(\Lambda\)CDM background
\citep{Linder2005}. Here
\(\Omega_m(a)=\Omega_{m0}a^{-3}H_0^2/H(a)^2\) is the matter density in
units of the critical density at epoch \(a\).

Let \(\mathbf{g}\) denote the peculiar gravitational acceleration, or force
per unit mass, generated by the density perturbations. Poisson's equation
and the growing-mode velocity--acceleration relation take the differential
form
\begin{equation}
    \begin{aligned}
    \nabla_{\mathbf{x}}\cdot\mathbf{g}
        &=-\frac{3}{2}aH^2\Omega_m\,\delta_m,\\
    \mathbf{u}&=\frac{2f}{3H\Omega_m}\,\mathbf{g},
    \end{aligned}
    \label{eq:linear_velocity_gravity}
\end{equation}
where \(\mathbf{u}\) is the physical peculiar velocity and the gravitational
field is curl-free. Thus, at a given epoch, the velocity is parallel and
proportional to the gravitational acceleration, with its amplitude set by
the expansion and growth rates. Taking the divergence gives the linear
continuity relation
\(\nabla_{\mathbf{x}}\cdot\mathbf{u}=-aHf\delta_m\).

For the linear forward model, we express this irrotational flow in terms of a
velocity potential \(\phi\):
\begin{equation}
    \mathbf{u}=-\nabla_{\mathbf{x}}\phi,
    \qquad
    \delta_m=\frac{\nabla_{\mathbf{x}}^2\phi}{aHf}.
    \label{eq:linear_velocity_potential}
\end{equation}
Density and velocity are therefore linear functions of the same underlying
field. These relations describe coherent motions on scales where nonlinear
evolution is weak.

We parameterize this forward model by the present-day potential
\(\phi_0\), related to the initial density by
\(\delta_{\rm init}=\nabla^2\phi_0/(H_0f_0)\). The spatially constant
potential mode is fixed since it affects neither density nor velocity.

At a fixed epoch, we adopt the linear bias relation
\(\delta_g=b\delta_m\). To obtain the redshift-space prediction, we
expand the common mapping in Eq.~(\ref{eq:forward_rsd_mapping}) to first
order in the density and velocity perturbations. Number conservation under
the radial shifts gives
\(\delta_g^s=\delta_g-(aH)^{-1}s^{-2}\partial_s(s^2v_r)\).
Substituting \(v_r=-\partial_s\phi\) and the density--potential relation
then gives \(\deltags=\mathcal L\phi\), where
\begin{equation}
    \mathcal L=
    \frac{b\nabla^2+
    f\left(\frac{2}{s}\frac{\partial}{\partial s}
    +\frac{\partial^2}{\partial s^2}\right)}{aHf}.
    \label{eq:linear_response}
\end{equation}
Here \(s=|\mathbf{s}|\) is the radial redshift-space coordinate. The
Laplacian term describes the biased real-space density. The radial derivative
terms describe the change in tracer density caused by peculiar velocities.
The \(2/s\) term retains the radial geometry of the observer-centered mapping.
Appendix~\ref{app:RSD} derives this response.

Writing the forward model in these potential variables, the fixed-epoch
prediction is \(\mathcal F_{\rm lin}[\phi]=M\mathcal L\phi\), with
the binary mask \(M\) defined above.

To extend this fixed-epoch response to the survey light cone, we scale
\(\phi_0\) to the redshift associated with each grid position using the
linear growth relation in Eq.~(\ref{eq:linear_density_growth}). In the
adopted implementation, the position-dependent scaling operator \(D_s\)
gives the response
\(\mathcal F_{\rm lin}[\phi_0]=M\mathcal L D_s\phi_0\), with the
background coefficients evaluated at the local redshift. We neglect the
velocity-induced change in the redshift used to evaluate the growth factors
and the additional displacement accumulated during the corresponding
light-travel-time shift. The growth scaling and full light-cone equation are
given in Appendix~\ref{app:linear_growth}.

\subsubsection{Zel'dovich dynamics and forward model}
\label{sec:zeldovich_dynamics}
\label{sec:nonlinear_reconstruction}
\label{sec:biasing}
\label{sec:multi_epoch}

Linear theory changes the amplitude of the density field at fixed comoving
positions. The Zel'dovich approximation instead follows the motion of
matter elements, retaining their displacement when computing the evolved
density \citep{Zeldovich1970}. A matter element initially at the Lagrangian
position \(\mathbf{q}\) has a displacement field
\(\boldsymbol{\Psi}(\mathbf{q})\) determined by the same initial density
used in Section~\ref{sec:linear_dynamics}:
\begin{equation}
    \nabla_{\mathbf{q}}\cdot\boldsymbol{\Psi}
    =-\delta_{\rm init},
    \qquad
    \nabla_{\mathbf{q}}\times\boldsymbol{\Psi}=0.
    \label{eq:za_displacement}
\end{equation}
Its comoving position at growth factor \(D\) is
\begin{equation}
    \mathbf{x}(\mathbf{q},D)
    =\mathbf{q}+D\,\boldsymbol{\Psi}(\mathbf{q}),
    \label{eq:za_position}
\end{equation}
and differentiation with respect to cosmic time gives its physical peculiar
velocity,
\begin{equation}
    \mathbf{v}(\mathbf{q},D)
    =a\dot D\,\boldsymbol{\Psi}(\mathbf{q})
    =aHfD\,\boldsymbol{\Psi}(\mathbf{q}).
    \label{eq:za_velocity}
\end{equation}
Here \(\mathbf{v}\) is the velocity of the matter element, corresponding
to the Eulerian velocity field \(\mathbf{u}\). The growth factor \(D\)
and rate \(f\) are the same as in linear theory.

Nonlinear density evolution arises from mass conservation at the displaced
positions: converging trajectories increase the density, while diverging
trajectories reduce it. Although the displacement is first order in
\(\delta_{\rm init}\), the density obtained from the changing volume of
a matter element is nonlinear. Expanding this volume change to first order
recovers \(\delta_m=D\delta_{\rm init}\). Retaining it captures the
transport of structure and anisotropic compression beyond linear theory.
Once trajectories cross, the approximation does not describe the subsequent
virialized motions.

Numerically, we evolve a uniform Lagrangian particle grid and obtain the
matter density by depositing the particles in real space
\citep{2026arXiv260608593N}.

To model the galaxy distribution, we assign tracer weights using the evolved
real-space matter density. A linear bias prescription can give negative
tracer densities in sufficiently underdense regions. We instead adopt the
normalized local relation \citep{2026arXiv260608593N}
\begin{equation}
    1+\delta_g(\mathbf{x})=
    \frac{[1+\delta_m(\mathbf{x})]^b}
    {\left\langle[1+\delta_m]^b\right\rangle}.
\end{equation}
This prescription preserves non-negative tracer densities and reduces to
linear bias for small fluctuations. Each particle receives a weight determined
by the matter density at its real-space position, as specified in
Appendix~\ref{app:zeldovich_details}.

We then use the particle's predicted radial velocity to shift its position
according to Eq.~(\ref{eq:forward_rsd_mapping}), carrying its tracer weight
with it. Depositing these weights at the shifted positions produces the
predicted redshift-space galaxy overdensity. We use triangular-shaped-cloud
(TSC) assignment, as specified in Section~\ref{sec:gridded_overdensity}.

We construct the light-cone prediction by evolving the same initial field
at a finite set of redshift nodes spanning the survey range. The effective
bias can vary between nodes because the luminosity distribution of a
flux-limited sample changes with distance
\citep{Norberg2001,Zehavi2011,Carrick2015,LilowNusser2021}. At each node,
we apply the bias weights and redshift-space mapping described above.
For each grid cell, we interpolate the two neighbouring epoch predictions
in the growth factor to obtain a single field
\(\delta_g^s(\mathbf{s}\mid\xi)\). The interpolation and normalization
are specified in Appendix~\ref{app:zeldovich_details}. The adopted nodes and
bias values are reported with the DESI results.

\subsection{Gaussian likelihood}
\label{sec:survey_response}
\label{sec:gaussian_likelihood}

Both reconstructions use a Gaussian likelihood for the observed overdensity
\(d\). At fixed cosmology, tracer bias, and survey response, its dependence
on the initial field enters through the forward prediction
\(\mathcal F[\delta_{\rm init}]\). Defining the cell residual
\(e_i=d_i-\mathcal F_i[\delta_{\rm init}]\), the likelihood used here is
\begin{equation}
    \begin{aligned}
    L(\delta_{\rm init})
        &\equiv p(d\mid\delta_{\rm init})\\
        &=p\!\left(d\mid\mathcal F[\delta_{\rm init}]\right)\\
        &=\prod_{i:M_i=1}\frac{1}{\sqrt{2\pi\sigma_{n,i}^2}}
        \exp\!\left[-\frac{e_i^2}{2\sigma_{n,i}^2}\right].
    \end{aligned}
    \label{eq:gaussian_likelihood}
\end{equation}
The product runs over cells retained by the survey mask \(M\), and
\(\sigma_{n,i}^2\) is the noise variance of the gridded overdensity in
cell \(i\). This expression adopts the diagonal covariance
\(N_{ij}=\delta^{\rm K}_{ij}\sigma_{n,i}^2\), where
\(\delta^{\rm K}_{ij}\) is the Kronecker delta, neglecting correlations
between cell noise contributions. In each reconstruction, \(N\) is
estimated from the random catalog and held fixed during inference.
The likelihood normalization is therefore independent of the trial initial
field. Equivalently,
\(-2\ln L=\mathbf e^\top N^{-1}\mathbf e+\mathrm{const}\), where
\(\mathbf e\) collects the residuals in the retained cells and \(N\) is
restricted to those cells.
Section~\ref{sec:recon_setup} specifies the catalog estimator and noise
weights. Appendix~\ref{app:bayesian_posterior} combines this likelihood
with the prior to give the posterior objective.

The Gaussian likelihood approximates the discrete sampling of the galaxy
field. The MDPL2 tests of \citet{2026arXiv260608593N} found similar MAP
velocities with Gaussian and Poisson likelihoods for the full halo sample,
with larger differences after dilution. This motivates the approximation
for well-sampled fields without implying equivalent posterior uncertainties.

\subsection{Posterior inference}
\label{sec:posterior_inference}

Combining the common prior and likelihood with each forward model gives
the posterior in Eq.~(\ref{eq:general_posterior}). We maximize it to obtain
the MAP initial field and its predicted density and velocity, referring
to the results as the ``linear reconstruction'' and ``Zel'dovich
reconstruction''. With the linear response, the posterior is Gaussian and
the MAP equals its mean. The nonlinear Zel'dovich response generally gives
a non-Gaussian posterior despite the Gaussian likelihood. We additionally
sample the linear posterior. The Zel'dovich inference is restricted to the
MAP solution.

\subsubsection{Linear MAP solution and posterior sampling}
\label{sec:linear_inference}

The Gaussian prior on \(\delta_{\rm init}\) induces a Gaussian prior on
\(\phi_0\), with covariance \(\xi_{\phi,0}\). For the linear light-cone
response of Section~\ref{sec:linear_reconstruction}, the MAP condition
reduces to a linear system.
Its solution is the Wiener-filter estimate \citep{1995ApJ...449..446Z}.
We solve it using preconditioned conjugate gradients
\citep{Hestenes1952MethodsOC}, evaluating derivatives and prior operations
with FFTs and applying the mask and noise weights in configuration space.
The posterior expressions and matrix-free implementation are given in
Appendix~\ref{app:linear_numerics}.

To sample the uncertainty around this solution, we draw an unconstrained
prior field, generate noisy mock data with the same forward response, and
reconstruct those mock data with the same solver. The difference between the
prior draw and its reconstruction is then added to the MAP field inferred
from DESI. This Hoffman--Ribak construction \citep{1991ApJ...380L...5H}
yields samples of the linear Gaussian posterior. Its explicit form is given in
Appendix~\ref{app:linear_posterior}.

\subsubsection{Zel'dovich MAP optimization}
\label{sec:zeldovich_inference}

For the Zel'dovich forward model, we optimize the whitened initial field
\(\xi\) using the common prior and likelihood. The objective is given in
Appendix~\ref{app:za_gaussian_objective}.

We minimize the objective using the limited-memory
Broyden--Fletcher--Goldfarb--Shanno (L-BFGS) algorithm, with gradients
computed by automatic differentiation in our \texttt{JAX}/\texttt{NumPyro}
implementation \citep{jax2018github,phan2019composable}.

\section{Numerical setup and catalog preparation}
\label{sec:recon_setup}

We now specify the grid used to implement the forward models,
and the construction of the data vector \(d\), survey mask \(M\), and
noise covariance \(N\) introduced in Section~\ref{sec:bayesian_framework}.
The detailed assignment formulas and shot-noise derivation are collected
in Appendix~\ref{app:gridding_noise}.

\subsection{Reconstruction grid}
\label{sec:grid}

We convert sky coordinates and observed redshifts to Cartesian redshift-space
positions using comoving distances evaluated with \texttt{Colossus}
\citep{2018ApJS..239...35D}. Both reconstructions use a cubic box of side
\(L_{\rm box}=3000\,h^{-1}\,\mathrm{Mpc}\), sampled by \(256^3\) cells
with spacing \(\Delta x\simeq11.7\,h^{-1}\,\mathrm{Mpc}\). The box encloses
the BGS sample with approximately \(183\,h^{-1}\,\mathrm{Mpc}\) of padding
between the outermost catalog coordinates and the box faces. The coordinate
conversion and box bounds are specified in Appendix~\ref{app:grid_geometry}.

The finite grid also sets the reference amplitude for comparisons with
reconstructed density fields. The RMS of a prior realization on this grid
is approximately \(0.71\), whereas \(\sigma_8=0.81\) measures fluctuations
smoothed with a spherical top-hat of radius \(8\,h^{-1}\,\mathrm{Mpc}\).
These are different measures of the same power spectrum. We use the
corresponding grid-level prior RMS when assessing reconstructed amplitudes.
Appendix~\ref{app:grid_prior_amplitude} describes the finite-mode restriction.

\subsection{Gridded overdensity and survey mask}
\label{sec:gridded_overdensity}

The data vector entering the likelihood is the observed galaxy density
contrast in redshift space, evaluated on a uniform grid. We construct it
by assigning the weighted BGS galaxies and random catalog to the grid,
with the randoms accounting for the survey selection. For the linear
reconstruction, we use cloud-in-cell (CIC) assignment. The linear forward
model acts directly on fields defined on this same fixed grid, so inference
does not require derivatives of the assignment kernel with respect to
particle positions.

For the Zel'dovich reconstruction, each trial initial density field (expressed in terms of the whitened variables)  determines
the displaced positions of model particles, whose deposition yields the
predicted density. MAP optimization therefore requires differentiating
through both the particle displacements and their assignment to the grid.
The CIC kernel is piecewise linear and has discontinuous first derivatives,
whereas TSC assignment has continuous first
derivatives with respect to particle position. We use TSC to provide smooth
deposition gradients in our \texttt{JAX}/\texttt{NumPyro} implementation.
We apply it to model-particle deposition in both real and redshift space,
and to the observed data and random catalogs, so that the predicted and
observed density fields use the same assignment kernel.

Let \(N_i\) be the weighted data count in cell \(i\), and    
\(\bar N_i\) the expected count inferred from the random catalog after
normalizing its total weight to that of the data. The observed data vector
used in the likelihood is
\begin{equation}
    d_i=\frac{N_i}{\bar N_i}-1.
    \label{eq:observed_overdensity}
\end{equation}
The random catalog supplies the spatial variation of the selection
function. Its normalization is a single global factor. The assignment
sums, weight normalization, and effective smoothing are given in
Appendix~\ref{app:weighted_overdensity}.

We determine survey support from the randoms independently of the observed
galaxy density, since an empty data cell may still lie within the survey.
For both reconstructions, the binary mask \(M\) retains cells with a
CIC-assigned weighted random count \(R_i^{\rm CIC}\geq1\). This threshold
excludes cells receiving only negligible weight from assignment across the
footprint boundary. We apply the same mask to the TSC-assigned data and
Zel'dovich prediction, so TSC's wider kernel does not enlarge the region
used in that likelihood.

\subsection{Noise covariance}

We estimate the shot-noise variance from the denser random catalog,
rescaled to the data normalization and accounting for the squared catalog
weights and assignment kernel. This avoids setting the likelihood weights
from fluctuations in the observed galaxy counts. The expected-count and
variance profiles are averaged radially before interpolation onto the grid,
as detailed in Appendix~\ref{app:weighted_shot_noise}.

These variances define the fixed diagonal noise covariance used in
Eq.~(\ref{eq:gaussian_likelihood}). Overlapping assignment kernels
correlate neighbouring cells, so neglecting these off-diagonal terms is
an approximation. The mask restricts the comparison to retained cells,
with this covariance supplying their noise weights.

% \subsection{Linear bias}

% For a fixed fiducial Planck-18 cosmology, the growth rate and Hubble parameter
% are specified, whereas the galaxy bias introduced in both forward models remains
% an effective modeling parameter. In the present context, this bias should not
% be interpreted as the physical bias of a single halo population, but rather as
% an effective survey-level bias that absorbs several ingredients of the observed
% sample, including the selection function, magnitude limits, radial evolution,
% and other observational systematics. Since the bias appears explicitly in the
% mean-field equation, it directly sets the amplitude of the reconstructed
% density field.

% To account for the redshift evolution of the observed tracer population while
% keeping the forward model simple, we parametrize the bias using two endpoint
% values, \(b_{\rm init}\) and \(b_{\rm final}\). In the linear reconstruction,
% the bias is varied smoothly along the line of sight across the reconstruction
% volume, interpolating between these two values. In the Zel'dovich
% reconstruction, the same prescription is applied across the sequence of
% realized epochs: the earliest epoch is assigned \(b_{\rm init}\), the latest
% epoch is assigned \(b_{\rm final}\), and the intermediate epochs are assigned
% linearly interpolated values. This provides a simple effective description of
% bias evolution without introducing a fully free redshift-dependent function
% \(b(z)\).

\section{Validation with MDPL2 mocks}
\label{sec:mock}

We apply the forward models and MAP inference of
Section~\ref{sec:bayesian_framework} to controlled mock catalogs for which the true peculiar velocities are known.
These tests assess the recovery of coherent radial flows and the effects of
survey boundaries and declining tracer density. The catalogs are designed
to isolate these effects, rather than reproduce the DESI observations.

\subsection{Mock construction}

We use the \(z=0\) halo catalog of the MultiDark Planck 2 (MDPL2)
simulation, obtained from the CosmoSim database
\citep{2016MNRAS.457.4340K}. MDPL2 evolves \(3840^3\) dark-matter particles
in a periodic \(1\,h^{-1}\,\mathrm{Gpc}\) cube with a Planck
\(\Lambda\)CDM cosmology. The catalog contains approximately
\(1.274\times10^8\) halos identified with the \textsc{Rockstar}
phase-space halo finder \citep{2013ApJ...762..109B}.

For these tests, we use the simulation volume with a \(128^3\) grid,
giving \(7.81\,h^{-1}\,\mathrm{Mpc}\) cells. The observer is at the box
center, and the halo positions are mapped to radial redshift space using
their simulated peculiar velocities and Eq.~(\ref{eq:forward_rsd_mapping}).
Only the \(z=0\) snapshot is used, so the forward models are evaluated at
a single epoch, without the light-cone evolution required for DESI. All
tracer weights are unity. Matching random catalogs supply the expected
counts, masks, and fixed likelihood noise weights using the construction
of Section~\ref{sec:recon_setup}.

We consider three configurations:
\begin{enumerate}
    \item A uniformly sampled catalog with full angular coverage,
    constructed by retaining \(20\%\) of the halos. Its mean occupancy is
    approximately 12 tracers per grid cell, with nearly uniform expected
    shot noise. This provides the baseline.
    \item The same sampling with a double-cone mask of half-opening angle
    \(40^\circ\) in each direction and an outer boundary at
    \(r=400\,h^{-1}\,\mathrm{Mpc}\). This configuration tests the combined
    effect of restricted angular coverage and a finite radial boundary.
    \item A catalog with full angular coverage and radial selection
    \begin{equation}
        S(r)=
        \begin{cases}
            \frac{1}{2}\exp(-r/r_0), & r\leq r_{\max},\\
            0, & r>r_{\max},
        \end{cases}
        \label{eq:mock_radial_selection}
    \end{equation}
    where \(r_0=100\,h^{-1}\,\mathrm{Mpc}\) and
    \(r_{\max}=400\,h^{-1}\,\mathrm{Mpc}\). Its density-noise amplitude
    averages approximately \(0.16\) and increases strongly with radius,
    weakening the likelihood constraint near the cutoff.
\end{enumerate}
The double-cone mask and declining radial selection are not applied
together in these tests.

We estimate the effective halo bias from
\(b(k)=\sqrt{P_h(k)/P_m(k)}\), where \(P_h\) is the shot-noise-corrected
halo power spectrum and \(P_m\) is the theoretical matter power spectrum.
The approximately constant large-scale value, \(b_{\rm eff}\simeq0.81\),
is used in the respective bias prescriptions of both forward models.
Each configuration is analyzed with the Gaussian prior and Gaussian
likelihood framework described above, and the diagnostics below use the
two resulting MAP velocity fields.

\subsection{Recovery of the radial velocity field}

We evaluate the predictions at \(20{,}000\) withheld halos within the
observed region and \(100<r<300\,h^{-1}\,\mathrm{Mpc}\). This radial
interval avoids the immediate vicinity of the observer and the outer
cutoff, although the double-cone sample still contains angular boundaries.
The linear velocity is evaluated on the reconstruction grid. For the
Zel'dovich prediction, particle velocities are deposited at their Eulerian
positions. In both cases, we interpolate the gridded field to the withheld
halo positions and project it along the line of sight.

\begin{figure*}[p]
    \centering
    \includegraphics[
        width=0.98\textwidth,
        height=0.86\textheight,
        keepaspectratio
    ]{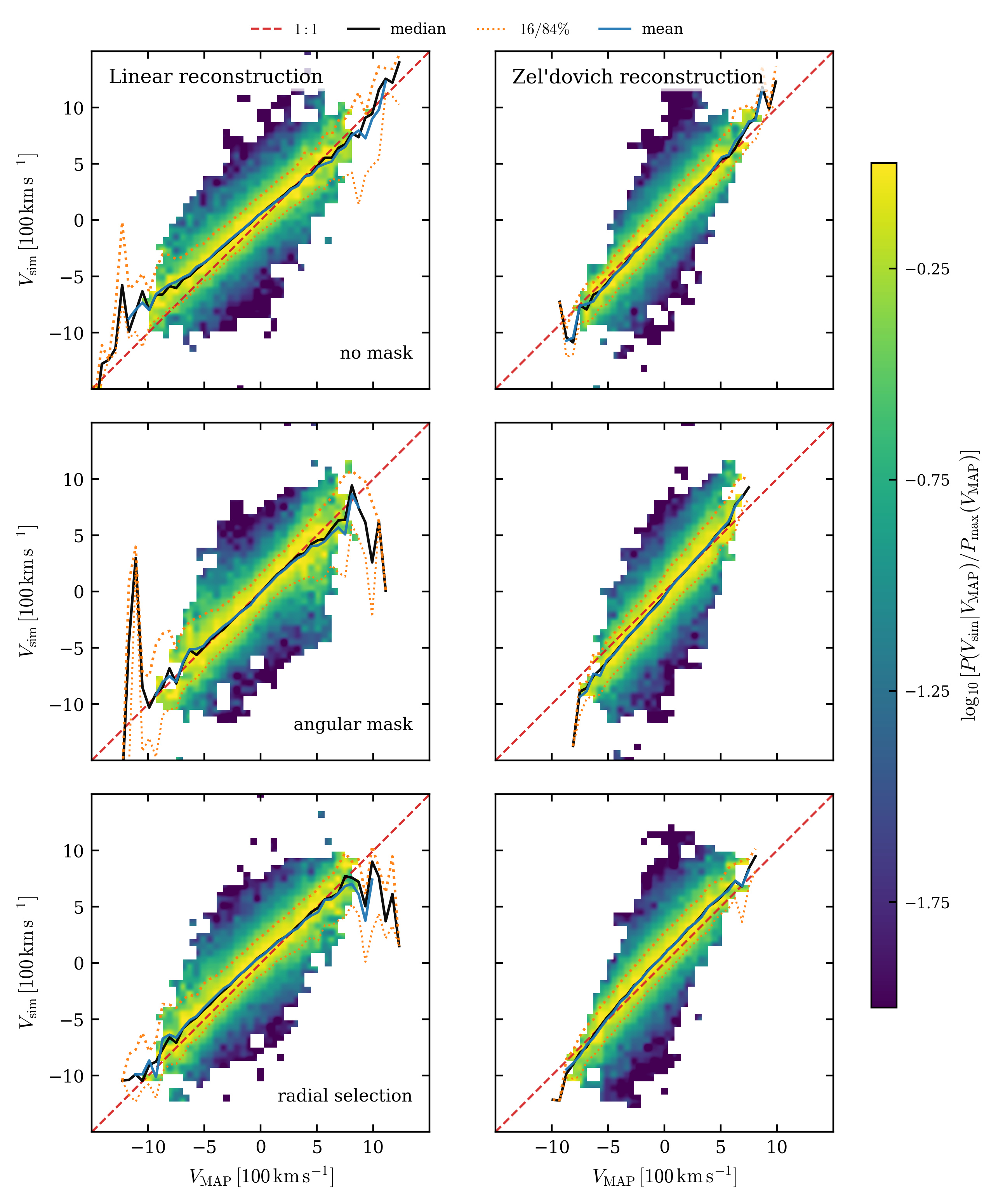}
    \caption{
    Conditional distributions of simulated radial velocity at fixed
    reconstructed MAP radial velocity for \(20{,}000\) withheld MDPL2 halos within
    \(100<r<300\,h^{-1}\,\mathrm{Mpc}\). The left and right columns show the linear
    and Zel'dovich reconstructions, respectively. Rows show the uniformly
    sampled full-sky catalog, the double-cone catalog, and the full-sky
    radial-selection catalog. Velocities are in units of
    \(100\,{\rm km\,s^{-1}}\). Sparsely populated tail bins should be
    interpreted cautiously.
    }
    \label{fig:mock_vel}
\end{figure*}

Figure~\ref{fig:mock_vel} shows the conditional distribution of the simulated
radial velocity \(V_{\rm sim}\) at fixed MAP prediction \(V_{\rm MAP}\).
The relation
\begin{equation}
    \left\langle V_{\rm sim}\mid V_{\rm MAP}\right\rangle
    \simeq V_{\rm MAP}
    \label{eq:mock_velocity_calibration}
\end{equation}
corresponds to an unbiased conditional mean and provides a diagnostic of
the predicted flow amplitudes. The scatter about this mean includes
reconstruction residuals, unresolved nonlinear motions, and interpolation
effects. It is not itself a posterior uncertainty estimate.

Within the displayed radial interval, the uniformly sampled full-sky
catalog follows the one-to-one relation closely. The double-cone
catalog shows a slight negative monopolar offset, while the full-sky
radial-selection catalog shows a somewhat larger positive offset. In
all three configurations, the Zel'dovich conditional mean lies closer to
the one-to-one relation than the linear result. This comparison concerns
the conditional mean under the tested sampling and resolution, rather
than establishing that one approximation is uniformly more accurate for
DESI.

Additional radial-shell comparisons, not shown here, reveal that these
offsets vary with radius. Near the observer, the monopolar component is
more pronounced in the double-cone catalog, whereas the offset in the
radial-selection catalog largely disappears where shot noise is low.
Near the outer boundary, both catalogs develop a large positive
monopole. The double-cone sample has a sharp geometric cutoff. The
radial-selection sample combines its cutoff with increasingly weak density
constraints. These tests therefore identify sensitivity to boundaries and
sampling, but do not separate their contributions to the outer-shell
offsets in the radially selected case.

\subsection{Implications and limits for DESI}

The tests demonstrate recovery of coherent radial flows in the well-sampled
interior of these simplified catalogs and show how systematic offsets can
develop near boundaries or where the sampling becomes sparse. They motivate
examining boundary dependence in the DESI results and using an intermediate,
well-sampled volume for the CF4 comparison. They do not determine a universal
distance from the boundary beyond which the DESI predictions are unbiased.

The mocks use a single simulation snapshot, a different grid resolution,
and idealized halo sampling. They do not reproduce the DESI light cone,
BGS tracer population, observational weights, or the joint angular and radial
selection. Consequently, neither the measured offsets nor the conditional
scatter can be transferred directly into error bars or a covariance for the
DESI velocity field or its derived statistics. The tests also do not assess
whether the uncertainty intervals from the linear posterior have the stated
coverage for DESI. Those intervals remain conditional on the adopted prior,
dynamics, bias, and noise model. No Zel'dovich posterior uncertainty is
estimated here.

Quantifying the DESI uncertainties requires an ensemble of realistic
light-cone mocks processed with the survey selection and the full inference
procedure. Such tests must include the combined effects of sampling,
boundaries, tracer bias, and observational corrections. Calibrating the
statistical significance of the CF4 comparison additionally requires its
distance-error model, which is absent from these simulation-velocity tests.

\section{Applications to DESI }
\label{sec:desi_results}

We apply the framework and numerical setup of
Sections~\ref{sec:bayesian_framework}--\ref{sec:recon_setup} in an
exploratory, preliminary application to DESI BGS. Our aim is to assess the
broad differences between the linear and Zel'dovich reconstructions,
including how each responds to survey boundaries, their overall agreement
with CF4 peculiar velocities, and the large-scale relative motions
quantified by the reflex dipole. After presenting the gridded data, we
examine the MAP fields and introduce the linear constrained realizations
(CRs) used in the subsequent flow and BAO diagnostics.

This application does not include a full assessment of statistical and
systematic uncertainties. The mock tests in Section~\ref{sec:mock} guide
the interpretation of survey effects but do not calibrate uncertainties
in the DESI results. We defer a comprehensive uncertainty analysis to
future work using more thoroughly validated survey catalogs and the next
DESI BGS data release. We also defer an assessment of the
luminosity-modulation method to later DESI releases, which will provide
the larger galaxy samples needed for that test.

\subsection{Gridded data and noise properties}

\begin{figure}[tbp]
\centering
\includegraphics[width=\columnwidth]
{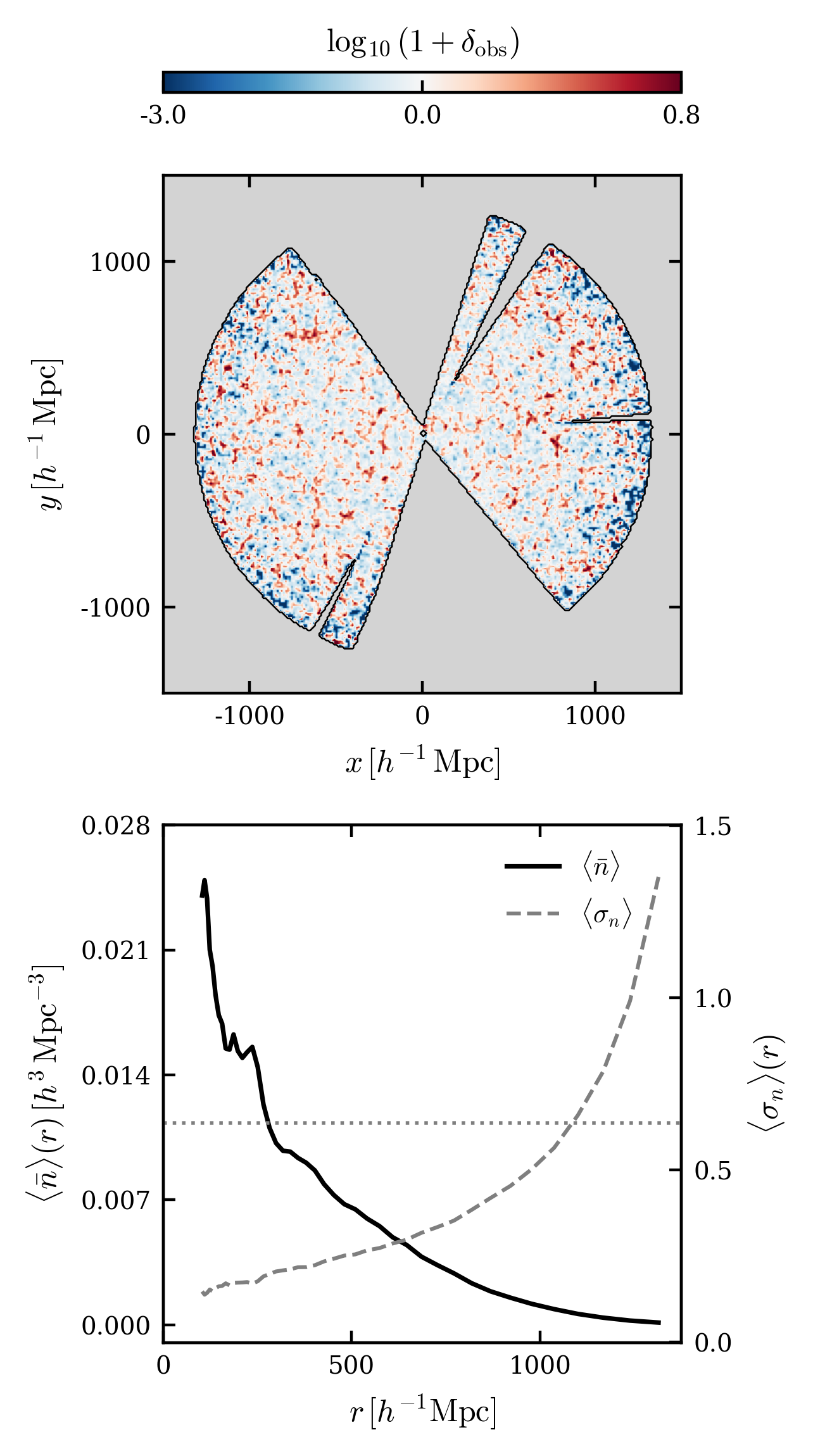}
\caption{Observed DESI DR1 BGS density field and radial noise properties. The upper
panel shows a slice through the observed redshift-space density contrast,
displayed as \(\log_{10}(1+\delta_{\rm obs})\), with the survey-mask
boundary overlaid in black. The lower panel shows the shell-averaged
expected tracer number density, \(\langle\bar n\rangle(r)\), as a solid
black curve, and the corresponding shell-averaged shot-noise amplitude
of the density contrast, \(\langle\sigma_n\rangle(r)\), as a dashed grey
curve. The horizontal dotted grey line marks the global mean shot-noise
amplitude over the observed region.}
\label{fig:DESI_data}
\end{figure}

Figure~\ref{fig:DESI_data} presents the gridded BGS data used for the
reconstruction. The upper panel shows a slice through the observed
redshift-space galaxy density contrast, \(d_i\equiv\delta_{{\rm obs},i}\),
constructed using Eq.~(\ref{eq:observed_overdensity}) on the same
\(256^3\) grid used to compare the data and forward-model predictions in
the likelihood, with cells of side length
\(\Delta x\simeq11.7\,h^{-1}\,\mathrm{Mpc}\). The increasingly
pronounced fluctuations between overdense and underdense cells at larger
distances reflect, in part, the rise in shot noise as the tracer number
density declines. The lower panel quantifies this radial variation through
shell averages of the expected tracer density,
\(\bar n_i=\bar N_i/(\Delta x)^3\), inferred from the random catalog,
and the dimensionless shot-noise amplitude,
\(\sigma_{n,i}=\sqrt{N_{ii}}\), entering the Gaussian likelihood in
Eq.~(\ref{eq:gaussian_likelihood}). The noise estimate accounts for the
catalog weights and assignment kernel, as described in
Appendix~\ref{app:weighted_shot_noise}.

Within the observed region, the unsmoothed gridded density contrast has a
mean of \(-0.012\) and an rms of \(1.3\). The small nonzero mean is
compatible with the estimator: the global random-catalog normalization
matches total weighted counts, whereas the mean quoted here averages the
cellwise ratios \(N_i/\bar N_i-1\) over the retained cells. With a
spatially varying selection function, this cell average is not constrained
to vanish.

The shell-averaged expected tracer density reaches approximately
\(0.024\,h^3\,\mathrm{Mpc}^{-3}\) in the innermost plotted shells and
decreases towards zero at \(r\simeq1300\,h^{-1}\,\mathrm{Mpc}\).
The corresponding shot-noise
amplitude rises from approximately \(0.2\) in the innermost shells to
\(1.3\) at large distances, with an average of approximately \(0.6\)
over the plotted shells. Through the likelihood weights, this radial trend
weakens the density constraint relative to the prior at larger distances.

\subsection{Reconstructed fields and survey-boundary effects}

For the linear forward model, we adopt a radially varying effective bias rather
than a constant value. The bias increases linearly from \(b=1\) at the observer
to \(b=2\) at a radial distance \(L_{\rm box}/2\), and is held fixed at
\(b=2\) beyond this distance. The midpoint of this adopted range is
\(1.5\), matching the constant effective bias used for the BGS BAO
reconstruction by \citet{Paillas2025}. This prescription phenomenologically
allows for the changing effective tracer population with distance and is not
intended as a measurement of the BGS galaxy bias.

For the Zel'dovich forward model, we use the same bias range while evaluating
the evolved fields at three redshift nodes, \(z=0\), \(0.3\), and \(0.6\). The
bias increases linearly from \(b=1\) at \(z=0\) to \(b=2\) at \(z=0.6\), and
the evolved fields are interpolated between these epochs to construct the
light-cone prediction, as described in Section~\ref{sec:multi_epoch}. These bias prescriptions are held
fixed during inference. The comparison therefore reflects both the choice
of dynamical approximation and the treatment of tracer bias.

The FFT representation imposes periodic boundary conditions on the full
\(3000\,h^{-1}\,\mathrm{Mpc}\) computational cube. The survey footprint
lies within this box, with approximately \(183\,h^{-1}\,\mathrm{Mpc}\)
of padding between the outermost catalog coordinates and the box faces
(Section~\ref{sec:grid}). The survey mask selects cells entering the
likelihood. It does not impose vanishing density or velocity at the
footprint boundary. This distinction is particularly relevant for velocities,
which depend on the density throughout the box, including regions outside
the observed footprint. The finite periodic representation also excludes
modes with wavelengths larger than the box size.

\begin{figure*}[tbp]
    \centering
    \includegraphics[width=1\textwidth]{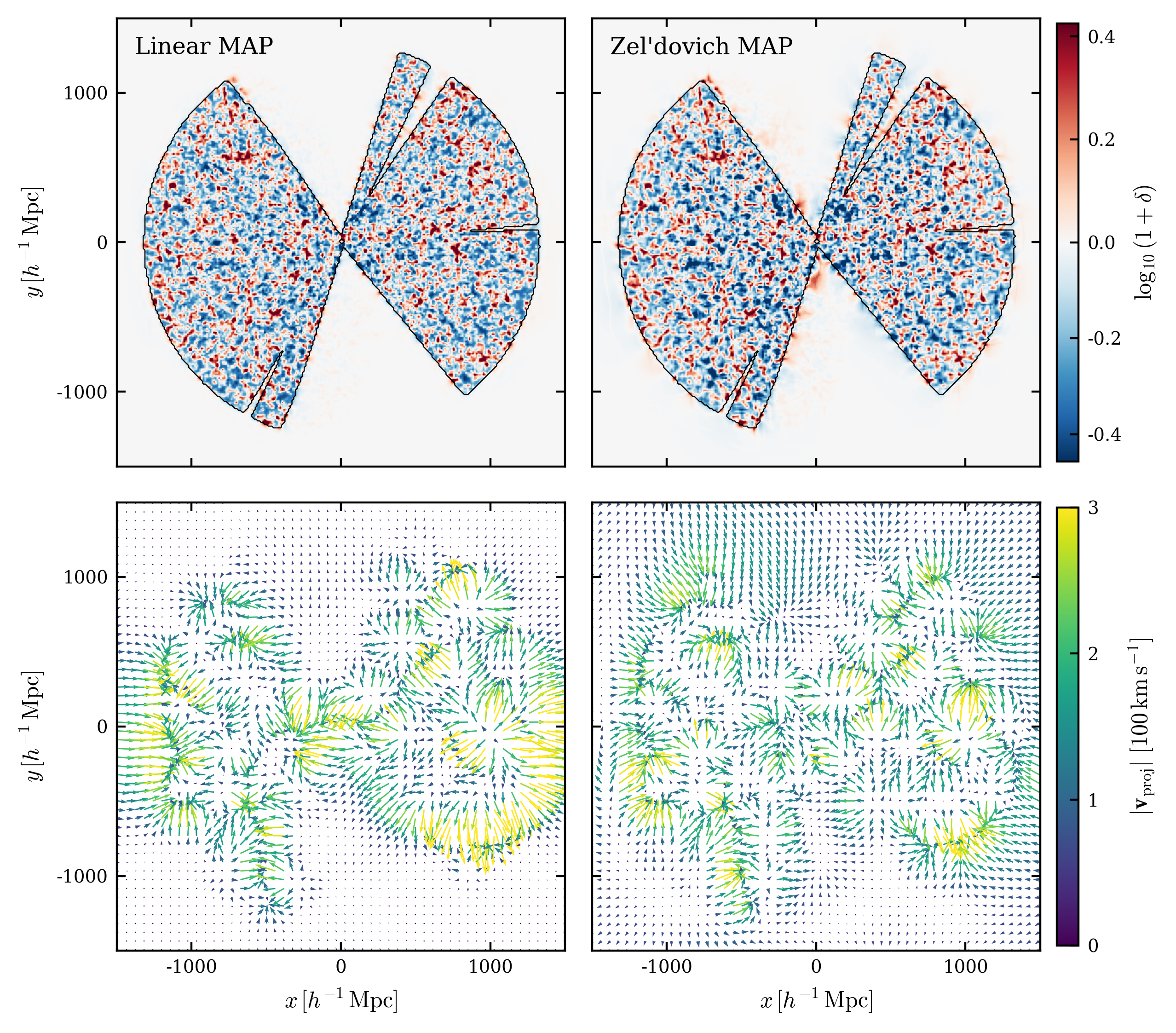}
    \caption{
    DESI DR1 BGS MAP density and velocity reconstructions obtained with the
    linear and Zel'dovich forward models. The density panels show the inferred
    large-scale structure relative to the survey footprint. The velocity panels
    are smoothed with a Gaussian kernel of width four grid cells
    (\(46.8\,h^{-1}\mathrm{Mpc}\)) for visualization.
    }
    \label{fig:DESI_density_velocity}
\end{figure*}

Figure~\ref{fig:DESI_density_velocity} shows the MAP matter-density and
velocity fields obtained with the linear and Zel'dovich forward models.
Both show similar large-scale density structures and velocity patterns
inside the footprint, with sharper density contrasts in the Zel'dovich
MAP. Visual inspection of the smoothed field residuals (not shown)
suggests closer agreement beyond roughly
\(50\,h^{-1}\,\mathrm{Mpc}\) inside the footprint boundaries.
Before the visualization smoothing specified in the caption, the velocity
rms at the \(11.7\,h^{-1}\,\mathrm{Mpc}\) grid resolution is approximately
\(400\,\mathrm{km\,s^{-1}}\) for both reconstructions inside the mask.

Near and beyond the survey boundary, the differences become more
pronounced. The linear velocity response is more localized on the inner
side of the mask and declines more rapidly outside it. The Zel'dovich
maps show density features and velocity patterns extending farther beyond
the footprint. To quantify these differences, we measure the density and
velocity rms against signed distance \(d_{\rm edge}\) from the mask:
positive distances are inside the survey and negative distances outside.

\begin{figure}[tbp]
    \centering
    \includegraphics[width=\columnwidth]{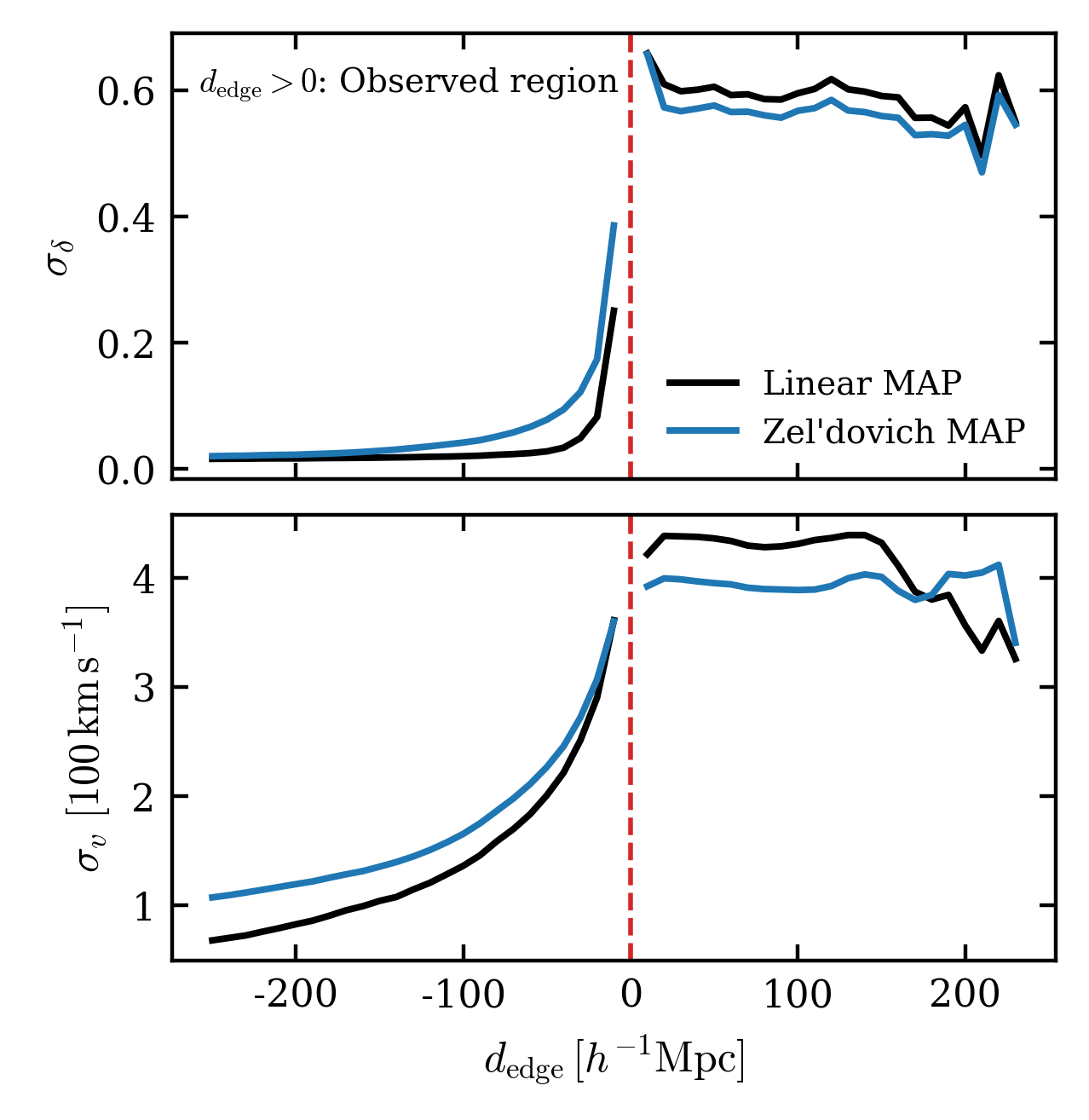}
    \caption{
    Boundary profiles of the reconstructed density and velocity fields. The
    upper panel shows the rms density contrast, \(\sigma_\delta\), as a
    function of signed distance from the survey mask boundary. The lower panel
    shows the corresponding rms velocity amplitude, \(\sigma_v\). Positive
    distances are inside the survey footprint, while negative distances are
    outside the footprint. Black and blue curves show the linear and
    Zel'dovich MAP fields, respectively. The red vertical dashed line marks
    the mask boundary.
    }
    \label{fig:boundary_profiles}
\end{figure}

The upper panel of Figure~\ref{fig:boundary_profiles} compares the density
rms of the Zel'dovich MAP (blue) and linear MAP (black). Outside the mask,
at \(d_{\rm edge}<0\) to the left of the red dashed line, the Zel'dovich
profile lies above the linear profile. Moving away from the boundary, the
linear density rms approaches its low exterior level sooner, whereas the
Zel'dovich profile retains elevated amplitudes farther outside the footprint.
The ordering reverses immediately inside the footprint, where the linear
density rms is slightly larger than the Zel'dovich value. The velocity-rms
profiles in the lower panel show the same general behaviour: the Zel'dovich
amplitude is larger outside the mask, whereas the linear amplitude is larger
inside the footprint in the vicinity of the boundary. At greater interior
distances, the velocity profiles approach and cross. These profiles measure
spatial field amplitudes, not reconstruction errors, and contain no
information about velocity direction. In particular, the declining MAP
density amplitude outside the footprint must not be confused with the prior
variance recovered by posterior samples.

A possible contribution to the more extended Zel'dovich profiles is the
transport of matter across the survey boundary. The likelihood constrains
the density after particle displacements, so particles initially outside
the footprint can contribute to the predicted density inside it.
Displacements that fit the interior can also generate structure outside
the mask, where no direct density constraint is applied. This mechanism
could leave a broader exterior imprint, but the rms profiles alone do not
establish that it causes the observed difference. Both reconstructions
couple observed and unobserved regions through prior correlations and the
nonlocal density--velocity relation. The different bias prescriptions and
assignment kernels further complicate attribution. Isolating the role of
particle transport would require controlled comparisons with matched bias
and gridding. We therefore use the footprint dependence to guide spatial
selections, without interpreting the difference between the MAP fields as
a reconstruction-error estimate.

\subsection{Linear posterior realizations}
\label{sec:linear_realizations}

We generate 15 linear constrained realizations with the Hoffman--Ribak
procedure of Section~\ref{sec:linear_inference}. These sample fluctuations
about the linear MAP under the adopted Gaussian posterior and are used
in the reflex-dipole and BAO comparisons below.

\begin{figure*}[tbp]
    \centering
    \includegraphics[width=1.0\textwidth]{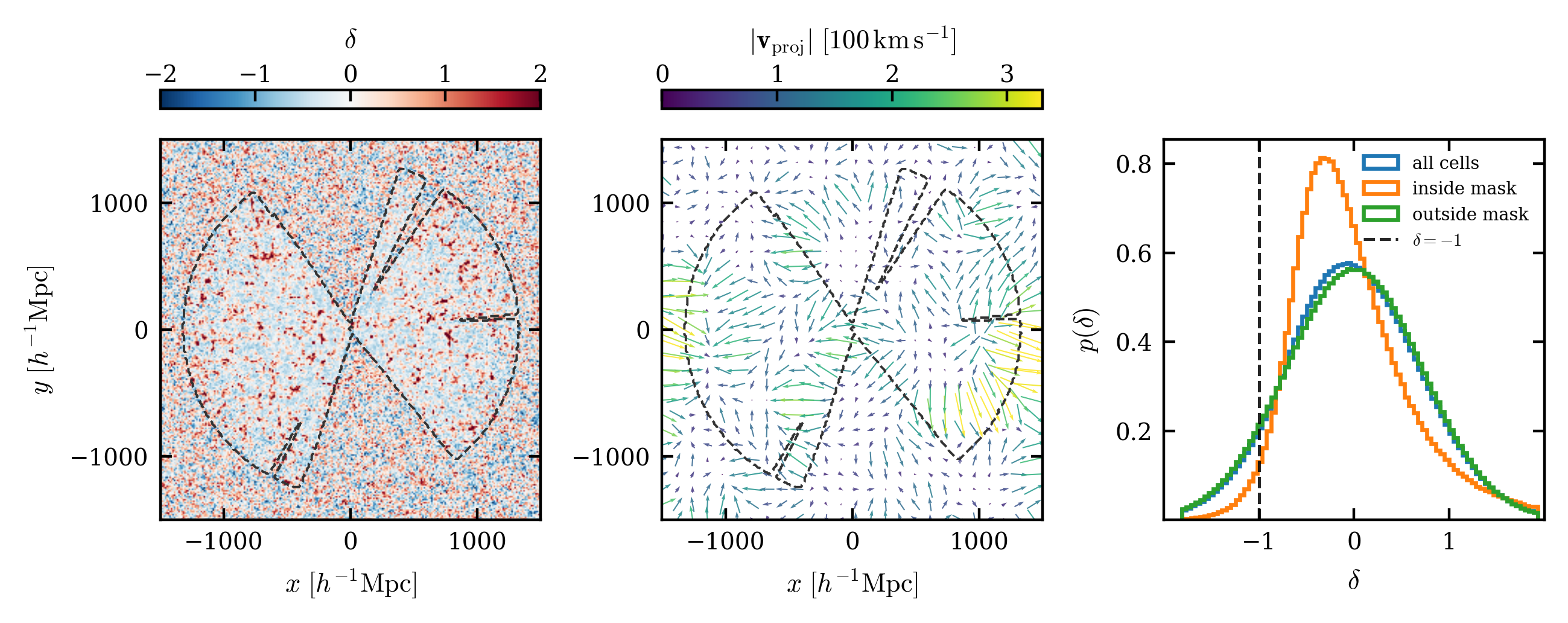}
    \caption{
        Example Hoffman--Ribak constrained realization. Left: density contrast
        in the reconstruction box. Middle: the corresponding projected
        velocity field, shown after Gaussian smoothing with a width of two grid
        cells for visual clarity. Right: one-point density distributions for
        cells inside and outside the observed survey region. The vertical
        dashed line marks \(\delta=-1\), below which the linear density field
        corresponds formally to a negative matter density.
    }
    \label{fig:cr_example}
\end{figure*}

The realization in Figure~\ref{fig:cr_example} retains the large-scale MAP
structures inside the footprint while adding fluctuations allowed by the
posterior. Outside, those fluctuations become larger as the direct data
constraints are lost. Its density rms is \(0.69\) inside and \(0.71\)
outside the mask, compared with \(0.62\) for the linear MAP inside. The
unconstrained prior realization used in this comparison has an rms of
approximately \(0.71\) at the same grid resolution.
% Numerical consistency check: the numerical setup quotes a grid-level prior RMS of
% 0.61. Verify the power-spectrum normalization, smoothing, and realization
% used for the 0.73 comparison before treating the two as the same reference.

Some cells have \(\delta<-1\), reflecting the absence of a positivity
constraint in the linear Gaussian density model. Averaging the 15
realizations gives density and velocity fields close to the MAP, without
an apparent systematic offset, providing a numerical check of the sampler.

These realizations describe uncertainty conditional on the assumed
cosmology, bias, linear dynamics, selection, and noise covariance. They do
not include uncertainty in those assumptions or calibrate the corresponding
DESI systematic errors. The small ensemble illustrates the posterior's
spatial behavior and its mean. It is not used here to establish a converged
covariance for the flow statistics, CF4 regression, or BAO measurement.

\subsection{Density-field diagnostics}

\begin{figure}[tbp]
    \centering
    \includegraphics[width=\columnwidth]{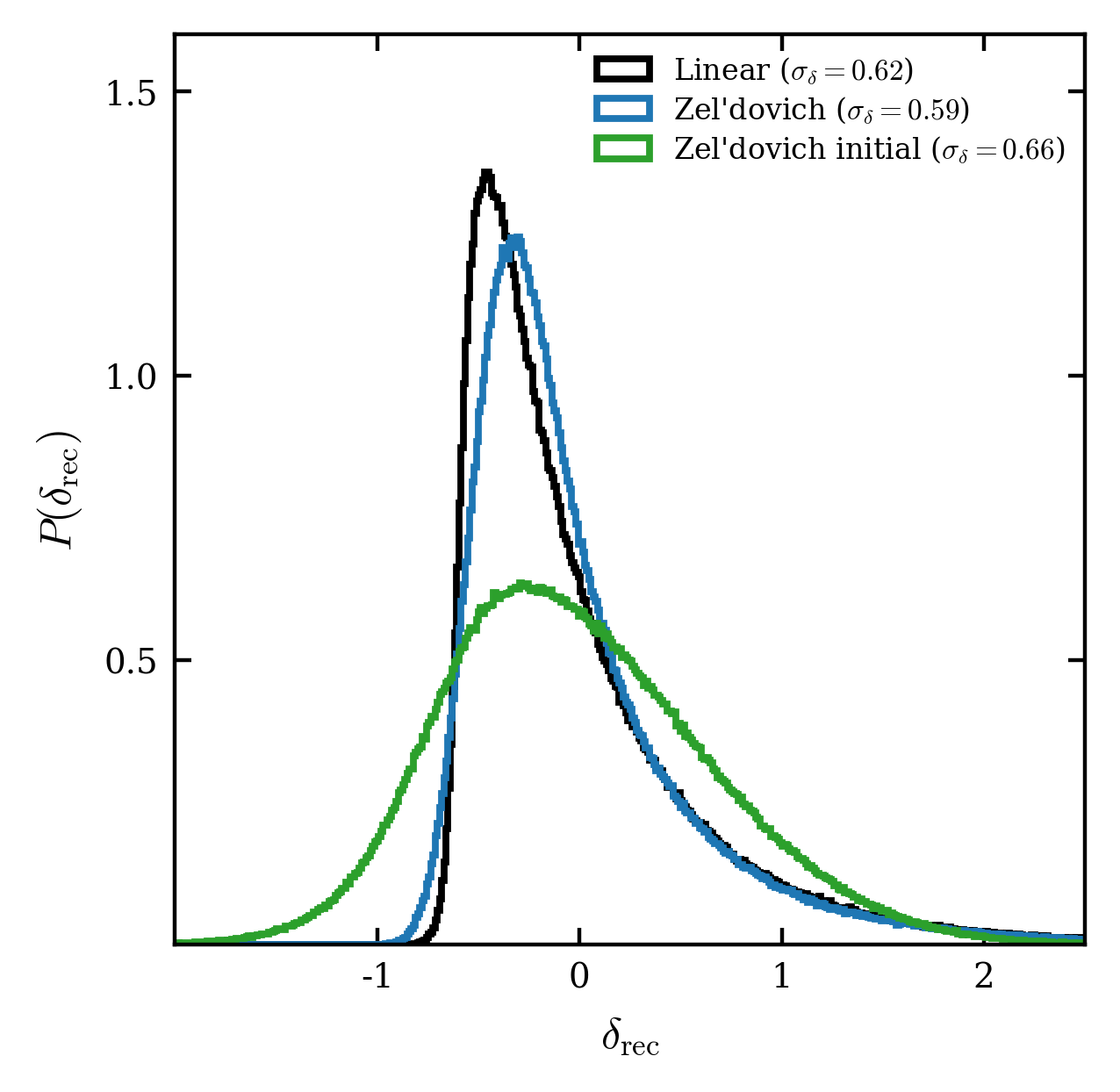}
    \caption{
    One-point distributions of the reconstructed real-space matter density
    within the survey footprint. Black and blue curves correspond to the
    linear and Zel'dovich MAP fields, respectively. The green curve is the
    inferred Zel'dovich initial density field, normalized to the present epoch.
    }
    \label{fig:density_pdfs}
\end{figure}

To compare the reconstructed density amplitudes, we examine the one-point
distributions within the survey footprint
(Figure~\ref{fig:density_pdfs}).

The rms density contrasts are \(0.62\) for the linear MAP, \(0.59\) for the Zel'dovich
MAP, and \(0.66\) for the inferred Zel'dovich initial field. The two MAP distributions have similar dispersions but differ in shape,
with the Zel'dovich reconstruction showing a less pronounced peak and a
more extended underdense tail, consistent with the sharper underdense
features in the reconstructed maps. These differences depend on the fixed
bias prescriptions and do not by themselves establish more accurate
density recovery.

\begin{figure*}[tbp]
    \centering
    \includegraphics[width=1.0\textwidth]{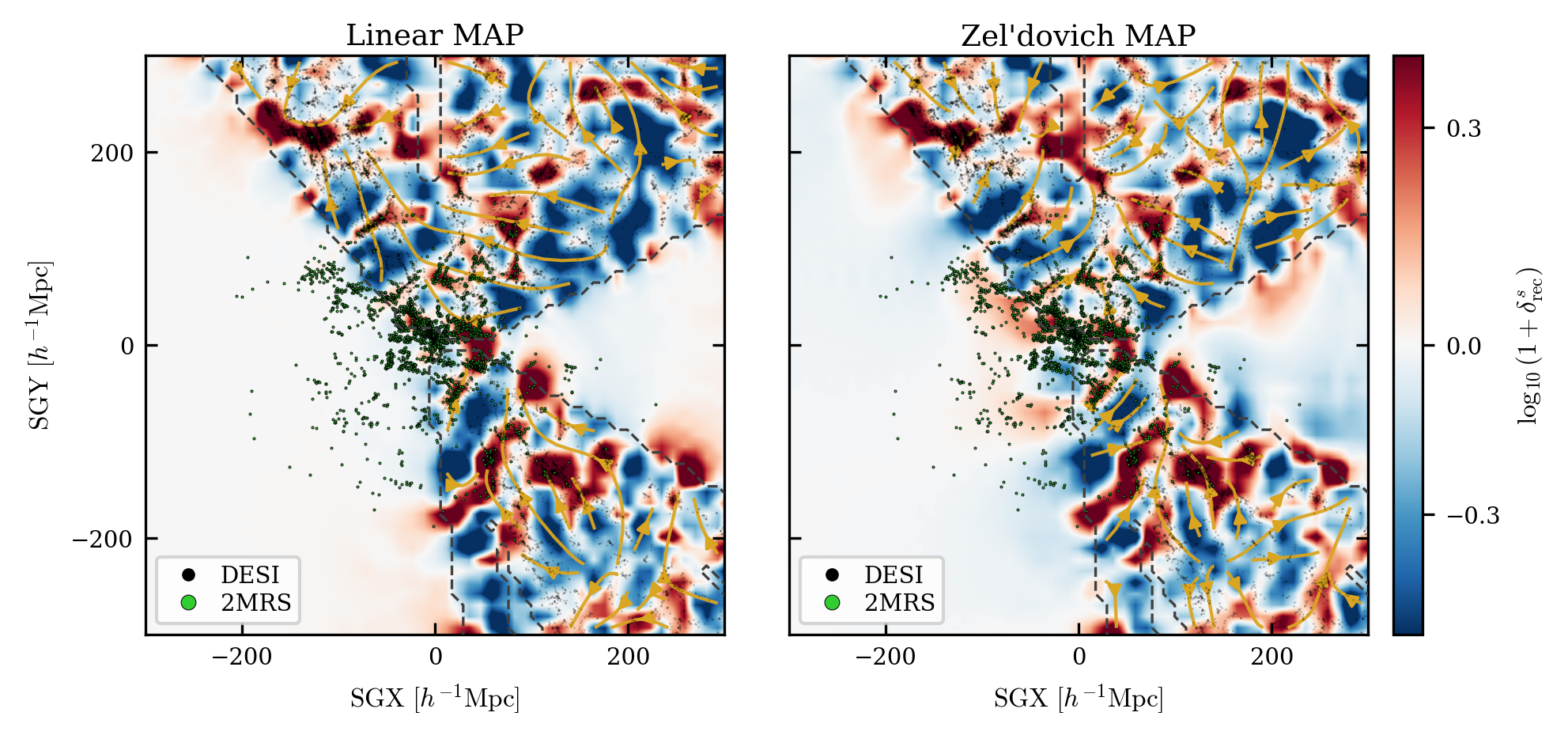}
    \caption{
        Redshift-space density reconstruction in the supergalactic plane. The color
        map shows the reconstructed density contrast in the SGX--SGY plane for the
        linear MAP reconstruction (left) and the Zel'dovich MAP reconstruction (right).
        Black points show DESI tracers, while green points show 2MRS galaxies in the
        same SGZ slab. The overlaid streamlines indicate the projected reconstructed
        velocity field after Gaussian smoothing with a width of two grid pixels, and are
        included to highlight the coherent large-scale flow pattern.
    }
    \label{fig:sgxy_density_streamlines}
\end{figure*}

Figure~\ref{fig:sgxy_density_streamlines} provides a spatial comparison in
the supergalactic SGX--SGY plane. Concentrations of DESI and 2MRS galaxies
generally coincide with high reconstructed density, while comparatively few
tracers occupy the underdense regions. Some density features near the mask
edge have few visible counterparts in the plotted slab. Finite slab
thickness and the support of the assignment kernels can affect this visual
comparison, so it does not identify those features as either physical
structures or reconstruction errors. The 2MRS overlay provides a qualitative
check of spatial correspondence. No statistical comparison is made here.

\subsection{Reflex dipole}

We define the reflex dipole as
\begin{equation}
    \mathbf{B}_{\rm shell}(r;\mathbf{x}_0)
    =
    \left\langle\mathbf{u}(\mathbf{x})-\mathbf{u}(\mathbf{x}_0)\right\rangle_
    {|\mathbf{x}-\mathbf{x}_0|\approx r},
\end{equation}
where \(\mathbf{x}_0\) is the shell center and the brackets denote an
average over the shell
\(|\mathbf{x}-\mathbf{x}_0|\in[r-\Delta r,r]\)
of width \(\Delta r\).
This quantity is of particular interest because, for a complete spherical
shell in linear theory, it depends only on the mass distribution enclosed
by the shell \citep{1994ApJ...421L...1N}.

We sample sphere centers randomly within the better-sampled survey region,
retaining only spheres fully contained within the footprint. We report the
mean amplitude over the \(N_{\rm c}(r)\) retained centers,
\begin{equation}
    B_{\rm shell}(r)
    =
    \frac{1}{N_{\rm c}(r)}
    \sum_{i=1}^{N_{\rm c}(r)}
    \left\|
    \mathbf{B}_{\rm shell}(r;\mathbf{x}_{0,i})
    \right\| .
\end{equation}

We use \(N_{\rm c}(r)=100\) for \(r<200\,h^{-1}\,\mathrm{Mpc}\).
At larger radii, fewer spheres fit within the footprint:
\(N_{\rm c}=23\) at \(r=200\,h^{-1}\,\mathrm{Mpc}\) and
\(N_{\rm c}=11\) at \(r=225\,h^{-1}\,\mathrm{Mpc}\).
No sampled sphere fits at \(r=250\,h^{-1}\,\mathrm{Mpc}\).

\begin{figure}[tbp]
    \centering
    \includegraphics[width=\columnwidth]{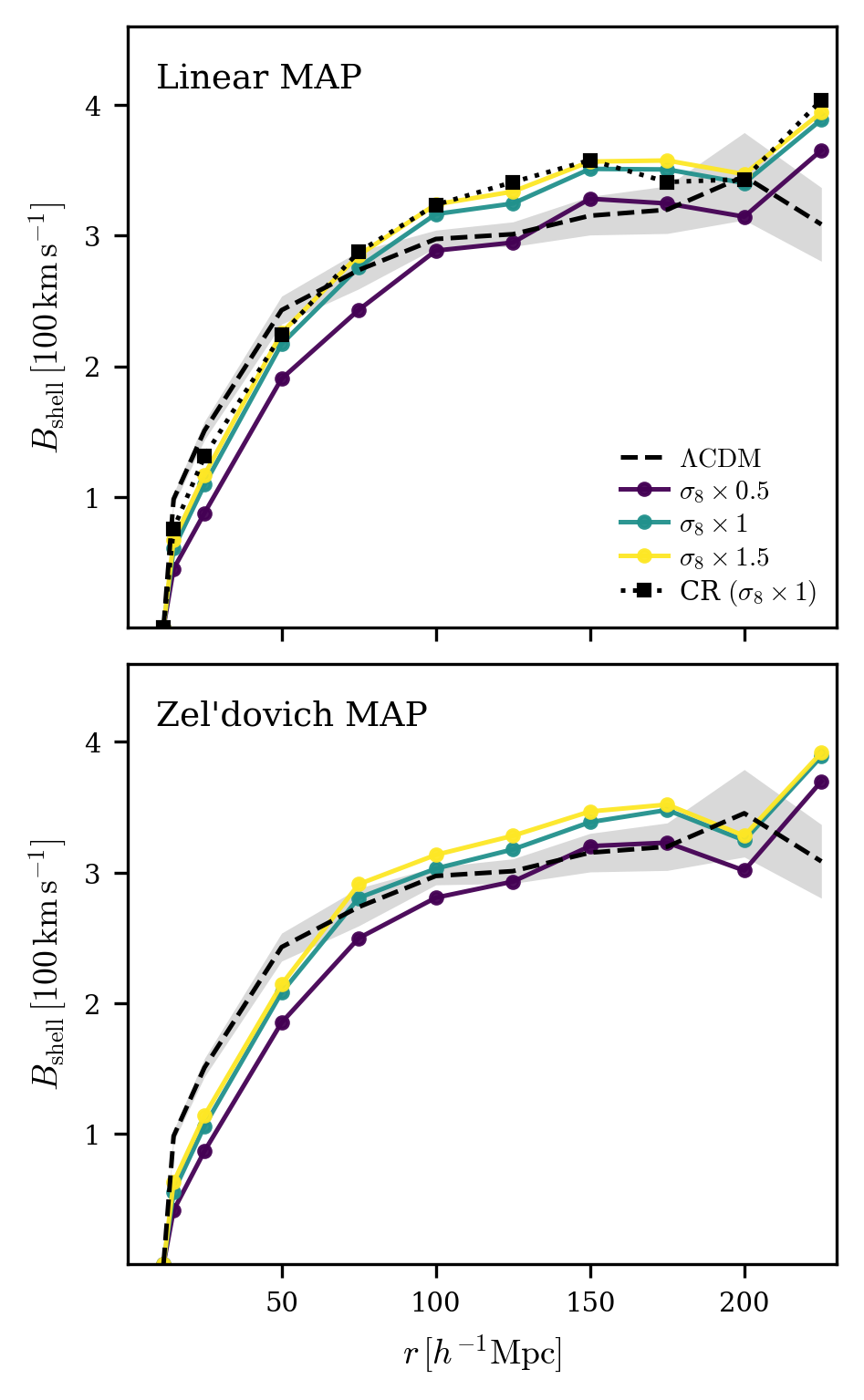}
    \caption{
    Mean reflex-dipole amplitude as a function of radius for the linear MAP
    (upper panel) and Zel'dovich MAP (lower panel). The purple, teal, and yellow
    curves correspond to \(\sigma_8/\sigma_8^{\rm fiducial}=0.5,\,1\), and
    \(1.5\), respectively. Amplitudes are averaged over spheres fully contained
    within the survey footprint, using shells whose width increases to
    \(25\,h^{-1}\,\mathrm{Mpc}\) at large radii.
    The black dashed curve and gray band show the mean and \(1\sigma\) scatter
    from random linear density and velocity fields generated using the
    fiducial \(\Lambda\)CDM power spectrum. In each realization, the
    reflex-dipole amplitude at a given radius is first averaged over the same
    number of spheres used for the DESI estimate at that radius. The band
    therefore describes the uncertainty in this sphere-averaged amplitude,
    not the dispersion of amplitudes among individual spheres of that radius.
    The black dotted curve with square
    markers in the upper panel shows one linear constrained realization at
    the fiducial value of \(\sigma_8\). No Zel'dovich CR is available.
    }
    \label{fig:shell_bulk}
\end{figure}

Figure~\ref{fig:shell_bulk} plots the reflex-dipole amplitude as a function
of radius for the linear and Zel'dovich MAP reconstructions, shown in the
upper and lower panels, respectively. The shell width increases with radius
to \(\Delta r=25\,h^{-1}\,\mathrm{Mpc}\) at large radii. The teal curves
correspond to the fiducial prior normalization. For comparison, the black
dashed curve gives the mean prediction obtained by generating random linear
density fields from the fiducial \(\Lambda\)CDM power spectrum, computing
their linear velocity fields, and applying the same reflex-dipole estimator.
Its gray band describes the uncertainty in the amplitude
averaged over the number of spheres used for the DESI estimate at each
radius, rather than the dispersion among individual spheres. The black
dotted curve in the upper panel shows one linear CR from
Section~\ref{sec:linear_realizations} at the fiducial prior normalization.

Both reconstructions show nearly identical radial trends: the amplitude is
close to zero at the smallest scale and rises to several hundred
\(\mathrm{km\,s^{-1}}\) near \(r=200\,h^{-1}\,\mathrm{Mpc}\).
This rise is consistent with the decreasing correlation between the central
and shell-averaged velocities. For a fixed prior normalization, the linear
and Zel'dovich MAP profiles closely agree over the full radial range, with no
consistent ordering between them.

At \(r<50\,h^{-1}\,\mathrm{Mpc}\), the reconstruction curves for both
dynamical approximations lie systematically below the fiducial linear
\(\Lambda\)CDM mean. This small-radius deficit persists across the tested
prior normalizations and is also present in the displayed linear CR. At
intermediate and larger separations, the fiducial MAP profiles generally lie
slightly above the theoretical mean. The origin and statistical significance
of these differences require further assessment.

For the linear reconstruction, the CR and fiducial MAP curves give similar
reflex-dipole amplitudes across the plotted radii. This comparison is
available only for the linear model, for which constrained realizations
have been generated.

The figure also shows reconstructions with the prior normalization reduced
to \(0.5\sigma_8^{\rm fiducial}\) (purple) and increased to
\(1.5\sigma_8^{\rm fiducial}\) (yellow). The modest changes in amplitude
across these choices indicate limited sensitivity to the prior normalization
within the adopted inference models. This spread is a sensitivity test,
not an uncertainty band or an independent test of the cosmological prior.

The reduced number of retained centers at the largest radii limits spatial
averaging, and the spheres need not be independent. The cancellation of exterior
contributions in the ideal linear limit also does not guarantee immunity
to boundary errors in the reconstructed fields.

\subsection{Comparison with CF4 peculiar velocities}
\label{sec:cf4_comparison}

We test the density--velocity connection predicted by gravitational
instability within the adopted \(\Lambda\)CDM model by comparing the DESI
MAP velocities with the independent CF4 group velocities. CF4 measurements
enter neither reconstruction likelihood. The comparison therefore confronts
the dynamical prediction from the DESI density field with distance-based
peculiar velocities, conditional on the assumed bias and survey model.

\begin{figure}[tbp]
    \centering
    \includegraphics[width=\columnwidth]{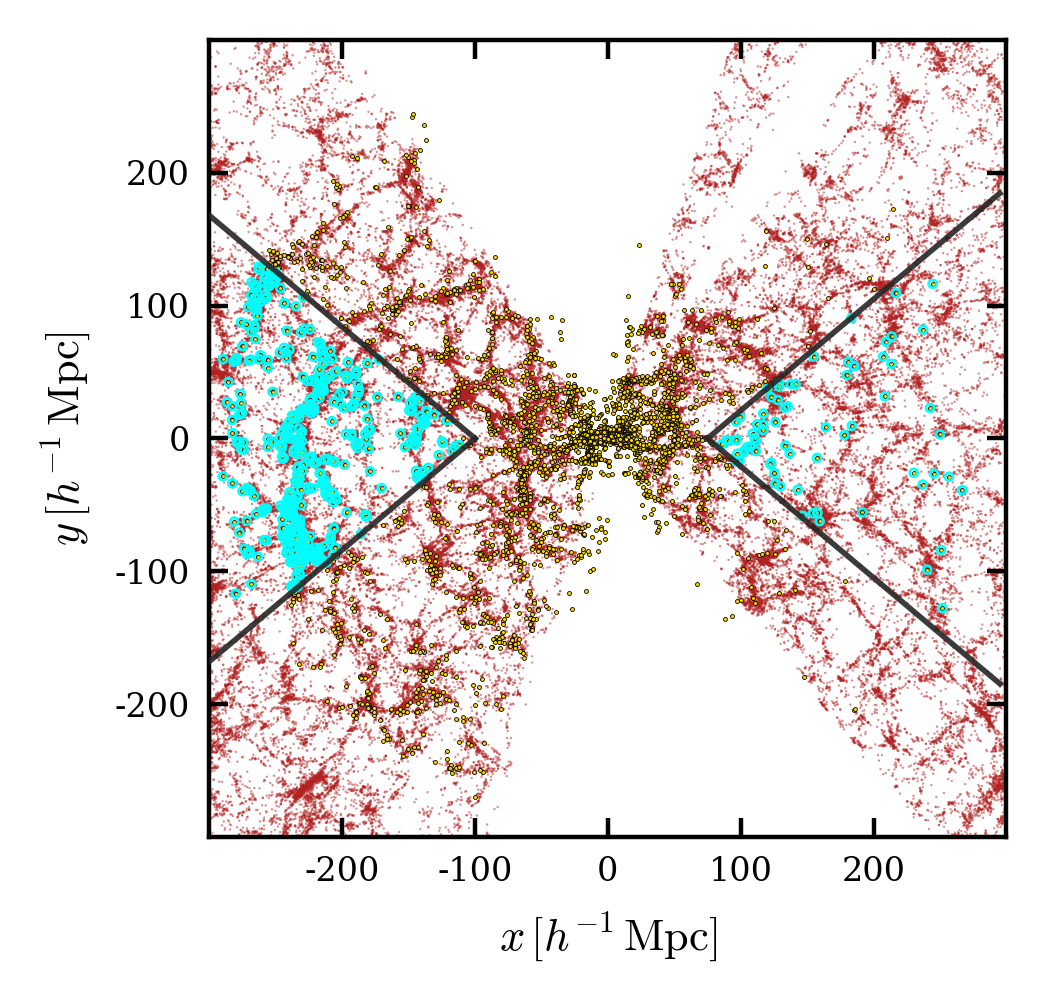}
    \caption{
        DESI and CF4 tracer distribution in the local comparison region. DESI
        tracers are shown in red, CF4 groups in dark yellow, and selected CF4
        groups inside the displayed slab in blue. The black lines show the
        projected double-cone boundaries used for the velocity comparison. The
        plot shows an \(xy\) slab approximately centered at \(z=0\), with a
        thickness of \(11.7\,h^{-1}\,\mathrm{Mpc}\).
    }
    \label{fig:desi_cf4_tracers}
\end{figure}

We choose an intermediate volume to reduce exposure to the immediate
observer region, the angular boundary, and the sparsely sampled outer
survey. This choice is motivated by the mock and DESI boundary diagnostics.
It is not a cut calibrated to guarantee a specified reconstruction error.
The selection comprises two oppositely directed cones inside the DESI
footprint, each with half-opening angle \(40^\circ\) and length
\(220\,h^{-1}\,\mathrm{Mpc}\). Their vertices are at
\(x=-100\,h^{-1}\,\mathrm{Mpc}\) and
\(x=75\,h^{-1}\,\mathrm{Mpc}\), extending towards negative and positive
\(x\), respectively. This selects \(3{,}605\) CF4 groups from the working
catalog described in Section~\ref{sec:data_cf4}. The velocity comparison
uses the full three-dimensional selection. Figure~\ref{fig:desi_cf4_tracers}
shows only a slab through it.

At each group's observed redshift-space position, we interpolate the MAP
velocity and project it along the line of sight. We then examine the
conditional mean CF4 velocity at fixed prediction, using the same direction
of conditioning as in Section~\ref{sec:mock}. If the residual has zero
conditional mean, the expected relation is
\begin{equation}
    \left\langle V_{\rm CF4}\mid V_{\rm MAP}\right\rangle
    \simeq V_{\rm MAP}.
\end{equation}
The residual includes distance errors, unresolved motions, and errors in
the prediction. Conditioning instead on the noisy CF4 velocity changes the
expected regression, as illustrated in Appendix~\ref{app:cf4_regression}.

\begin{figure*}[tbp]
    \centering
    \includegraphics[width=1.0\textwidth]{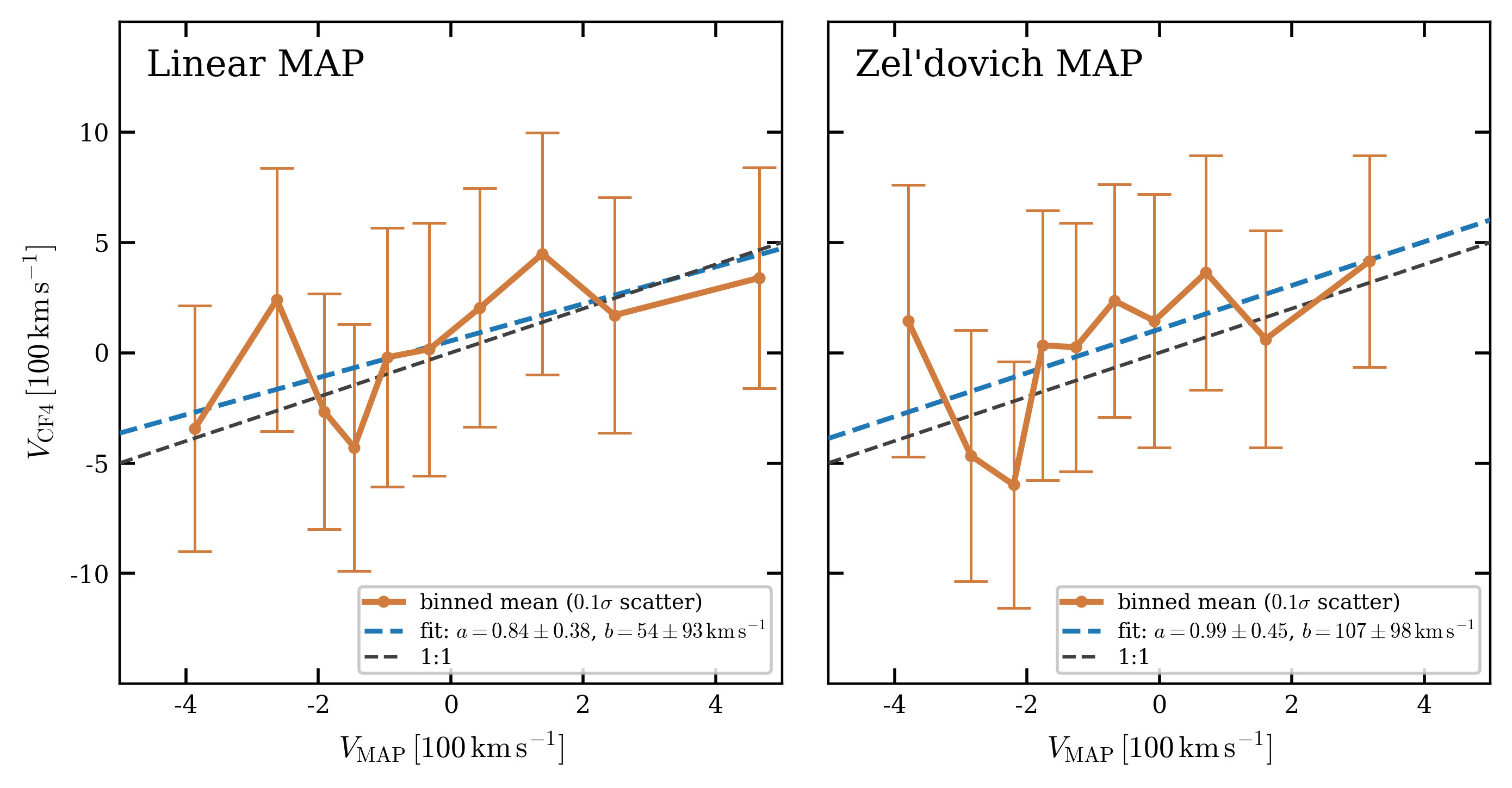}
    \caption{
        CF4 velocity comparison in the selected double-cone region. The left
        and right panels show the linear and Zel'dovich MAP reconstructions,
        respectively. The horizontal axis gives the line-of-sight MAP velocity,
        and the vertical axis the observed CF4 radial peculiar velocity from the
        \texttt{Vpds} field. The orange curves show the binned conditional means,
        with error bars indicating \(0.1\sigma\) of the within-bin scatter among
        the CF4 groups. The dashed blue lines show the best-fitting regressions,
        while the dashed black lines mark the one-to-one relation.
    }
    \label{fig:cf4}
\end{figure*}

For both reconstructions, the binned conditional mean follows the
one-to-one trend. Writing the fitted relation as
\(V_{\rm CF4}=aV_{\rm MAP}+b\), the linear MAP gives
\(a=0.84\pm0.38\) and
\(b=54\pm93\,\mathrm{km\,s^{-1}}\), while the Zel'dovich MAP gives
\(a=0.99\pm0.45\) and
\(b=107\pm98\,\mathrm{km\,s^{-1}}\).
Both slopes are consistent with unity, and neither intercept shows a
significant departure from zero given the quoted fit uncertainties.
Although the Zel'dovich slope is closer to unity, the uncertainties do not
establish a statistically significant improvement over the linear MAP.
The average
binned \(1\sigma\) scatter among individual group velocities is
approximately \(5500\,\mathrm{km\,s^{-1}}\). This broad distribution
includes the large distance-indicator uncertainties as well as unresolved
velocity structure and reconstruction residuals.

The agreement of the conditional means supports the density--velocity
connection predicted by the adopted cosmology and gravitational-instability
framework in this selected volume. Its statistical precision is not given
by the error bars, which describe individual velocities rather
than errors on the means. The quoted regression uncertainties quantify the
precision of the fitted slopes and intercepts under the fitting assumptions.
Consistency with a zero intercept does not establish the absence of a
systematic monopole. A full assessment of the comparison must also account
for the CF4 distance-error model, spatial correlations, and uncertainty in
the reconstructed predictor. The idealized mocks do not provide that
calibration.

\subsection{Exploratory BAO reconstruction}

We apply the inferred displacement fields to reduce the large-scale motions
that broaden the BAO feature. The DESI DR1 procedure described by
\citet{Paillas2025} estimates a first-order (Zel'dovich) displacement by
solving the linear continuity equation for a Gaussian-smoothed galaxy-density
field, including linear bias and redshift-space distortions. Its iterative
FFT solver refines this linear solution, after which galaxies and randoms
are shifted to reduce BAO broadening. This use of Zel'dovich displacements
does not involve inferring initial conditions through forward particle
evolution.

Our linear BAO correction uses the same linear density--displacement
relation, applied to the Bayesian inferred matter field. Its prior
regularization, bias prescription, and finite target redshift differ from
the DESI implementation. In our Zel'dovich case, the initial field is
inferred by displacing particles from a uniform Lagrangian grid and
matching their evolved density to the observations. This retains the
nonlinear density response to the displacements within the likelihood.
The inferred particle displacements are then deposited on the redshift-space
grid for the BAO correction.

In both cases, we apply the same redshift-dependent shift to the DESI data
and random catalogs toward a target epoch \(z_{\rm target}=8\), then
measure the correlation function. Shifting the randoms accounts for the
change in the survey reference density. This is a displacement correction,
not a full dynamical reversal to the target epoch. The catalog-level
estimator, displacement equations, and shift-to-target construction are
described in Appendix~\ref{app:bao_details}.

\begin{figure}[tbp]
    \centering
    \includegraphics[width=\columnwidth]{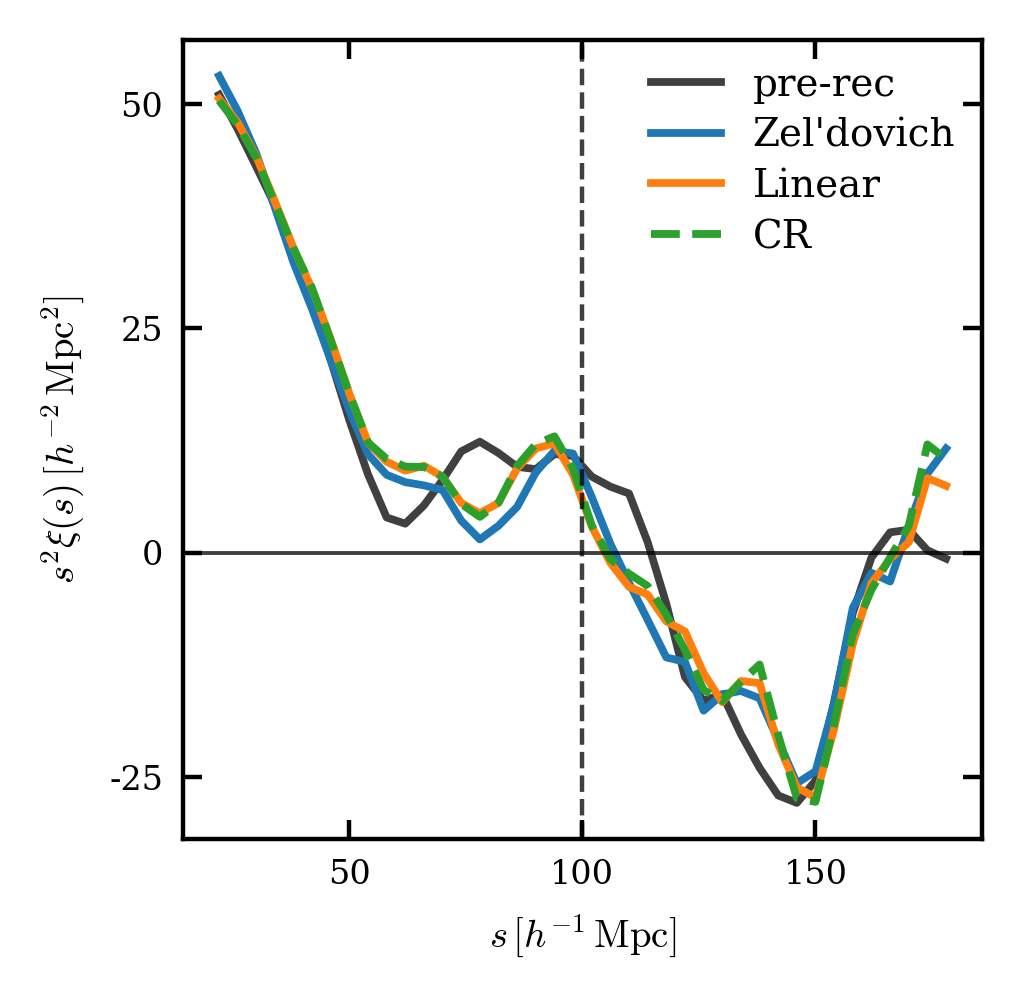}
    \caption{Preliminary BAO reconstruction test. We show the monopole correlation function
    before reconstruction, after applying displacements from the linear and
    Zel'dovich MAP fields, and after applying displacements from one example linear
    constrained realization.}
    \label{fig:bao}
\end{figure}

Figure~\ref{fig:bao} shows a more distinct BAO feature after either shift.
The Zel'dovich result has a sharper peak relative to its surroundings,
whereas the linear result is broader and peaks at a somewhat smaller
separation. This visual comparison suggests differences in how the two
inferred displacement fields remove BAO broadening. It does not establish
an improvement in BAO-scale accuracy or precision: no template fit,
correlation-function covariance, or complete systematic-error assessment
is included in this exploratory test.

\section{Discussion}
\label{sec:discussion}

The linear and Zel'dovich reconstructions yield broadly similar large-scale
density fluctuations and coherent velocity patterns inside the DESI DR1 BGS
footprint. The Zel'dovich MAP has sharper density features and retains
larger density and velocity amplitudes outside the observed region.
Particle displacements across the mask could contribute to this extended
response. The bias prescriptions also differ, so the comparison does not
fully isolate the effect of nonlinear dynamics.

The MDPL2 tests recover coherent radial flows, with conditional means closer
to the one-to-one relation for Zel'dovich in the configurations tested.
Their sensitivity to boundaries and sparse sampling motivates our selection
of an intermediate volume for the CF4 comparison. However, these idealized
mocks do not reproduce the DESI light cone, tracer population, or survey
selection and cannot calibrate DESI uncertainties or guarantee unbiased
predictions within the selected volume.

The CF4 comparison in Figure~\ref{fig:cf4} tests the density--velocity
connection using independent distance measurements that enter neither
reconstruction likelihood. Both reconstructions give slopes close to unity,
although the uncertainties are large and do not establish a significant
difference in calibration. Large individual distance errors limit the
precision of this comparison. A more precise test requires propagating
distance errors, spatial correlations, and uncertainty in the reconstructed
predictor.

The reflex-dipole profiles recovered by the linear and Zel'dovich methods
have nearly identical radial trends and amplitudes, with no consistent
ordering between them and only modest sensitivity to the tested prior
normalizations. At \(r<50\,h^{-1}\,\mathrm{Mpc}\), all reconstruction
curves lie below the reference mean from random linear density fields
generated with the fiducial \(\Lambda\)CDM power spectrum, whereas at
intermediate and larger separations the fiducial MAP profiles generally lie
slightly above it. The origin and significance of these differences require
further assessment of reconstruction response, tracer bias, sampling, and
nonlinear dynamics. At large radii, the small number of retained spheres
limits spatial averaging. Varying the prior normalization does not estimate
these uncertainties. The reflex dipole also differs from a conventional
bulk flow, so these results do not resolve claims of bulk-flow anomalies on
much larger scales.
%
%The constrained realizations provide a first assessment of field
%uncertainty within the linear model. The displayed linear CR and MAP yield
%similar reflex-dipole amplitudes, while the ensemble of 15 realizations
%illustrates the larger field fluctuations allowed outside the footprint.
%This ensemble does not provide a converged covariance for the derived flow
%statistics, and no Zel'dovich posterior has been sampled. The posterior
%uncertainty is also conditional on fixed cosmology, bias, dynamics,
%selection, and noise covariance. Neither it nor the differences between the
%MAP fields account for all reconstruction systematics.

Both reconstructions sharpen the BAO feature, with a more pronounced peak
for Zel'dovich. Determining whether this improves BAO-scale accuracy or
precision requires template fits, covariance estimates, and realistic mocks.

The DESI Peculiar Velocity Survey offers a substantially larger comparison
sample through the Fundamental Plane (FP) for early-type galaxies and the
optical Tully--Fisher (TF) relation for spirals. The reported DR1 samples
contain about \(100{,}000\) FP and \(10{,}000\) TF galaxies before
analysis-specific cuts
\citep{2025arXiv251203226R,2025arXiv251203227D}. These peculiar-velocity
catalogs are awaiting public release, motivating our use of CF4.

For independent measurements with comparable errors and a similar spread
in predicted velocities, the larger DR1 sample could reduce the linear-MAP
slope uncertainty to about \(0.07\), roughly a factor of five improvement.
The original full-survey forecast of about \(200{,}000\) FP and TF
measurements \citep{2025MNRAS.539.3627S} would give about \(0.05\) under
the same assumptions. These illustrative estimates depend on the usable
count after spatial cuts and grouping. Calibration errors, correlations,
and reconstruction uncertainties limit the gains, and larger samples need
not reduce individual-velocity scatter or systematic offsets.

The present analysis assesses reconstruction behavior with cosmology and
tracer bias held fixed. We do not infer cosmological parameters, including
the growth rate, or quantify statistical concordance with \(\Lambda\)CDM.
We defer these goals to planned work with DESI DR2, including reassessment
of the small-radius reflex-dipole deficit and velocity comparisons using
both FP and TF distances. This will require realistic DESI light-cone
mocks processed through the full inference procedure, propagation of
reconstruction and distance uncertainties, and joint inference of tracer
bias with the underlying field.

Kinematic Sunyaev--Zel'dovich detections using galaxy samples selected from
DESI imaging \citep{2025JCAP...05..057M,2025arXiv250621684L} motivate
another test. Cross-correlating CMB temperatures with our reconstructed
radial velocities could test their coherence with the ionized-gas motions,
subject to modeling the gas distribution.

A further promising avenue uses the spatial modulation of redshift-based
galaxy luminosities by peculiar velocities. With approximately half a
million SDSS DR7 galaxies, \citet{Feix2014} constrained bulk flows and higher
velocity moments at \(z\sim0.1\). These results motivate extending the
method to the larger galaxy samples expected from later DESI releases as
a complementary test of density-predicted velocities. We have not tested
this approach here. Its application requires modeling luminosity evolution
and selection and controlling spatial variations in photometric calibration.

The reconstructed fields are available from the authors upon request;
requests for reconstructions with alternative cosmological parameters
and tracer-bias prescriptions are also welcome.

\begin{acknowledgments}
This research was supported by the Israel Science Foundation.

This research used data obtained with the Dark Energy Spectroscopic Instrument (DESI). DESI construction and operations is managed by the Lawrence Berkeley National Laboratory. This material is based upon work supported by the U.S. Department of Energy, Office of Science, Office of High-Energy Physics, under Contract No. DE–AC02–05CH11231, and by the National Energy Research Scientific Computing Center, a DOE Office of Science User Facility under the same contract. Additional support for DESI was provided by the U.S. National Science Foundation (NSF), Division of Astronomical Sciences under Contract No. AST-0950945 to the NSF’s National Optical-Infrared Astronomy Research Laboratory; the Science and Technology Facilities Council of the United Kingdom; the Gordon and Betty Moore Foundation; the Heising-Simons Foundation; the French Alternative Energies and Atomic Energy Commission (CEA); the National Council of Humanities, Science and Technology of Mexico (CONAHCYT); the Ministry of Science and Innovation of Spain (MICINN), and by the DESI Member Institutions: www.desi.lbl.gov/collaborating-institutions. The DESI collaboration is honored to be permitted to conduct scientific research on I’oligam Du’ag (Kitt Peak), a mountain with particular significance to the Tohono O’odham Nation. Any opinions, findings, and conclusions or recommendations expressed in this material are those of the author(s) and do not necessarily reflect the views of the U.S. National Science Foundation, the U.S. Department of Energy, or any of the listed funding agencies.

The CosmoSim database used in this work is a service of the
Leibniz Institute for Astrophysics Potsdam. The MultiDark database was
developed in cooperation with the Spanish MultiDark Consolider Project
CSD2009-00064. We gratefully acknowledge the Gauss Centre for
Supercomputing and the Partnership for Advanced Supercomputing in
Europe for supporting the MultiDark simulation project through
computing time on the SuperMUC supercomputer at the Leibniz
Supercomputing Centre.

OpenAI ChatGPT, including its Codex coding environment, was used under
author supervision to improve the clarity and organization of the
manuscript and to assist with the refinement of analysis and
visualization code. All AI-assisted text and code were reviewed, tested,
and revised by the authors, who take full responsibility for the final
manuscript and results.
\end{acknowledgments}

% \begin{contribution}
% \end{contribution}

\software{
Astropy \citep{2013A&A...558A..33A,2018AJ....156..123A,2022ApJ...935..167A},
CAMB \citep{2000ApJ...538..473L,2026arXiv260714854L},
Colossus \citep{2018ApJS..239...35D},
JAX \citep{jax2018github},
NumPyro \citep{phan2019composable},
NumPy, SciPy, Matplotlib, pycorr, Corrfunc
}
          
\appendix
\renewcommand{\theHequation}{\Alph{section}.\arabic{equation}}
\section{Posterior objective and linear solver}
\label{app:bayesian_posterior}

With residual vector \(\mathbf e=d-\mathcal F[\delta_{\rm init}]\) and fixed covariances,
the posterior of Eq.~(\ref{eq:general_posterior}) is
\begin{equation}
    -\ln p(\delta_{\rm init}\mid d)
    =\frac{1}{2}\mathbf e^\top N^{-1}\mathbf e
    +\frac{1}{2}\delta_{\rm init}^\top C_{\rm init}^{-1}\delta_{\rm init}
    +\mathrm{const}.
    \label{eq:general_gaussian_posterior}
\end{equation}
The MAP minimizes this expression. In whitened variables, the prior term
is \(\xi^\top\xi/2\).

\subsection{Linear response and solution}
\label{app:linear_numerics}
\label{app:RSD}
\label{app:linear_growth}

Number conservation under Eq.~(\ref{eq:forward_rsd_mapping}) gives
\((1+\delta_g^s)s^2\,\mathrm{d}s=(1+\delta_g)r^2\,\mathrm{d}r\).
Here \(r=|\mathbf{x}|\) is the real-space radius. Expanding to first order yields
\(\delta_g^s=\delta_g-(aHr^2)^{-1}\partial_r(r^2v_r)\).
Substituting \(v_r=-\partial_r\phi\) and
\(\delta_g=b\nabla^2\phi/(aHf)\), and replacing \(r\) by \(s\) at
this order, gives Eq.~(\ref{eq:linear_response}).

The potential growth factor and adopted light-cone response are
\begin{equation}
    D_\phi(z)=\frac{a(z)H(z)f(z)}{H_0f_0}D(z),
    \qquad \mathcal R=M\mathcal L D_s,
\end{equation}
where \(D_s\) applies \(D_\phi\) at each grid position and the coefficients
of \(\mathcal L\) are evaluated at the local epoch. The present-day MAP
potential satisfies
\begin{equation}
    [\xi_{\phi,0}^{-1}+\mathcal R^\top N^{-1}\mathcal R]\bar\phi_0
    =\mathcal R^\top N^{-1}d.
    \label{eq:matvec}
\end{equation}
The posterior covariance is
\(C_{\rm post}=(\xi_{\phi,0}^{-1}+\mathcal R^\top N^{-1}\mathcal R)^{-1}\).
We solve the system by preconditioned conjugate gradients. Cartesian
derivatives and prior operations use FFTs. Radial contractions, mask,
and noise weights act in configuration space. The adjoint reverses this
sequence, avoiding explicit dense matrices.

\subsection{Posterior sampling}
\label{app:linear_posterior}

Draw \(\phi_0^{\rm u}\sim\mathcal N(0,\xi_{\phi,0})\), generate
\(d^{\rm u}=\mathcal R\phi_0^{\rm u}+\eta\) with
\(\eta\sim\mathcal N(0,N)\), and solve Eq.~(\ref{eq:matvec}) for its
mean field \(\bar\phi_0^{\rm u}\). The constrained realization
\begin{equation}
    \hat\phi_0=\bar\phi_0+(\phi_0^{\rm u}-\bar\phi_0^{\rm u})
\end{equation}
has ensemble mean \(\bar\phi_0\) and covariance \(C_{\rm post}\).

\section{Catalog gridding and noise}
\label{app:gridding_noise}

\subsection{Grid and count estimator}
\label{app:grid_geometry}
\label{app:weighted_overdensity}
\label{app:grid_prior_amplitude}

Distances are computed from \(\chi(z)=c\int_0^z\mathrm{d}z'/H(z')\).
The BGS Cartesian coordinates span
\(-1316.738\) to \(1316.670\,h^{-1}\,\mathrm{Mpc}\), within the
observer-centered box of Section~\ref{sec:recon_setup}. For its \(256^3\)
grid, \(k_{\rm Nyq}=\pi/\Delta x\simeq0.268\,h\,\mathrm{Mpc}^{-1}\)
in each Cartesian direction. The grid RMS uses only the supported Fourier
modes and the adopted smoothing, so it differs from the top-hat statistic
\(\sigma_8\).

For normalized assignment kernels \(\sum_iK_i(\mathbf{s})=1\),
\begin{equation}
    N_i=\sum_{j\in\mathrm{data}}w_jK_i(\mathbf{s}_j),
    \qquad
    R_i=\sum_{p\in\mathrm{rand}}w_pK_i(\mathbf{s}_p).
\end{equation}
The expected count is \(\bar N_i=\alpha R_i\), with global normalization
\(\alpha=\sum_{\rm data}w_j/\sum_{\rm rand}w_p\), giving
\(d_i=N_i/\bar N_i-1\). CIC and TSC conserve total weight. Their smoothing
windows are retained to avoid amplifying noise and aliasing near the
Nyquist frequency.

\subsection{Noise weights}
\label{app:weighted_shot_noise}

For a Poisson tracer density \(n(\mathbf{s})\), weighted assignment gives
\begin{equation}
    \operatorname{Cov}(N_i,N_j)
    =\int\mathrm{d}^3s\,n(\mathbf{s})w^2(\mathbf{s})
    K_i(\mathbf{s})K_j(\mathbf{s}).
\end{equation}
We retain only its diagonal and estimate it from the random catalog:
\begin{equation}
    \sigma_{n,i}^2\simeq
    \frac{\alpha\sum_{p\in\mathrm{rand}}w_p^2K_i^2(\mathbf{s}_p)}
         {\bar N_i^2}.
    \label{eq:shot_noise_variance}
\end{equation}
Expected counts and variances are averaged in eight radial shells and
interpolated onto the Cartesian grid. The resulting covariance is held
fixed, with off-diagonal assignment correlations neglected.

\section{Zel'dovich implementation}
\label{app:zeldovich_details}
\label{app:za_gaussian_objective}

The normalized bias prescription of Section~\ref{sec:biasing} is implemented
by assigning each evolved particle the real-space weight
\begin{equation}
    W_{\rm bias}(\mathbf{x})=
    \frac{[1+\delta_m(\mathbf{x})]^{b-1}}
         {\langle[1+\delta_m]^b\rangle}.
\end{equation}
After the redshift-space shift, the tracer density in cell \(j\) is
\(\rho_{g,j}=\sum_pW_{\rm bias}(\mathbf{x}_p)K_j(\mathbf{s}_p)\),
where \(K_j\) is the TSC kernel specified in
Section~\ref{sec:gridded_overdensity}.
Bias normalization is imposed separately at each epoch. Between adjacent
nodes \(z_\alpha\) and \(z_{\alpha+1}\), the light-cone field is
\begin{equation}
    \begin{aligned}
    \delta_g^s(\mathbf{s})&=(1-t)\delta_g^{s,\alpha}(\mathbf{s})
        +t\delta_g^{s,\alpha+1}(\mathbf{s}),\\
    t&=\frac{D[z(\mathbf{s})]-D(z_\alpha)}
             {D(z_{\alpha+1})-D(z_\alpha)}.
    \end{aligned}
\end{equation}

For the whitened variables, Eq.~(\ref{eq:general_gaussian_posterior}) becomes
\begin{equation}
    -\ln p(\xi\mid d)=\frac{1}{2}\xi^\top\xi+
    \frac{1}{2}\sum_{j\in\mathrm{mask}}
    \frac{[d_j-\delta_{g,j}^s(\xi)]^2}{\sigma_{n,j}^2}
    +\mathrm{const}.
    \label{eq:za_negative_log_posterior}
\end{equation}
In the independent unit-weight count approximation,
\(\sigma_{n,j}^{-2}=\bar N_j\). Catalog weights and assignment modify
this as described in Appendix~\ref{app:weighted_shot_noise}.

\section{Velocity comparison and BAO shifts}

\subsection{CF4 regression convention}
\label{app:cf4_regression}

For zero-mean jointly Gaussian variables, write
\(V_{\rm CF4}=V_{\rm MAP}+\eta\), with \(\eta\) uncorrelated with
\(V_{\rm MAP}\). Reversing the conditioning then gives
\begin{equation}
    \langle V_{\rm MAP}\mid V_{\rm CF4}\rangle=
    \frac{\sigma_{\rm MAP}^2}{\sigma_{\rm MAP}^2+\sigma_\eta^2}
    V_{\rm CF4}.
\end{equation}
This illustrates regression attenuation. It does not propagate uncertainty
in the reconstructed predictor.

\subsection{BAO displacement implementation}
\label{app:bao_details}

For the linear MAP field,
\(\boldsymbol{\Psi}_{\rm lin}(\mathbf{k})=
 i\mathbf{k}\delta_{\rm lin}(\mathbf{k})/k^2\), with the zero mode fixed.
For Zel'dovich, we attach the displacement of Eq.~(\ref{eq:za_displacement})
to each particle and deposit it at its predicted redshift-space position.
Combining Eqs.~(\ref{eq:za_position})--(\ref{eq:za_velocity}) with
Eq.~(\ref{eq:forward_rsd_mapping}) gives the position at each epoch and hence
\begin{equation}
    \Delta\mathbf{s}(z)=\mathbf{s}(z_{\rm target})-\mathbf{s}(z).
\end{equation}
Shifts are interpolated between redshift nodes and applied identically to
data and random tracers. At field level, the shifted contrasts give
\(\delta_{\rm BAO}=\delta_d-\delta_s\). At catalog level, the
Landy--Szalay-type estimator uses the shifted randoms to account for the
displaced survey window and selection function.

%% For this sample we use BibTeX plus aasjournalv7.bst to generate the

\bibliography{velocity_reconstruction}{}
\bibliographystyle{aasjournalv7}



\end{document}